\documentclass[11pt]{article}
\usepackage{amsmath, latexsym, amssymb, amscd, amsfonts, amsthm,wrapfig}
\usepackage[center,small]{caption} 
\numberwithin{equation}{section}
\usepackage{graphicx,epsfig}

\newcommand{\F}{{\cal F}}
\newcommand{\RR}{\mathbb{R}}
\newcommand{\CC}{\mathbb{C}}
\newcommand{\CS}{{\cal S}}
\newcommand{\inp}[2]{\langle #1,#2 \rangle}

\newcommand{\ket}[1]{|#1 \rangle}
\newcommand{\bra}[1]{\langle #1 |}
\newcommand{\ketbra}[1]{|#1 \rangle\langle #1|}
\newcommand{\half}{{\textstyle \frac{1}{2}}}
\newcommand{\dt}{\textstyle \frac{d}{dt}}

\newtheorem{theorem}{Theorem}[section]

\newtheorem{lemma}[theorem]{Lemma}

\begin{document}

\title{Optimal estimation of qubit states with
continuous time measurements}
\author{M\u{a}d\u{a}lin Gu\c{t}\u{a}$^{1}$, Bas Janssens$^{1}$
and Jonas Kahn$^{2}$\\\\
%$^{1}$University of Nijmegen, Department of Mathematics\\
% Toernooiveld 1, Postbus 9010, 6500 GL Nijmegen, The Netherlands\\\\
$^{1}$University of Nottingham, School of Mathematical Sciences,\\
University Park, Nottingham NG7 2RD, UK\\\\
%$^{1}$University of Nijmegen, Department of Mathematics\\
% Toernooiveld 1, Postbus 9010, 6500 GL Nijmegen, The Netherlands\\\\
$^{2}$Universit\' e Paris-Sud 11, D\' epartement de Math\' ematiques, \\
B\^{a}t 425, 91405 Orsay Cedex, France
}
\date{}
\maketitle

\abstract{
\setlength{\parindent}{0pt}
\setlength{\parskip}{1mm}
\noindent 
We propose an adaptive, two steps strategy, for the estimation of mixed qubit 
states. We show that the strategy is optimal in a local minimax sense for the 
trace norm distance as well as other locally quadratic figures of merit. Local minimax optimality means that given $n$ identical qubits, there exists no estimator which can perform better than the proposed estimator on a neighborhood of size $n^{-1/2}$ of an arbitrary state. In particular, it is asymptotically Bayesian optimal for a large class of prior distributions.

We present a physical implementation of the optimal estimation strategy based on continuous time measurements in a field that couples with the qubits.

The crucial ingredient of the result is the concept of local asymptotic normality (or LAN) for qubits. This means that, for large $n$, the statistical model described by
$n$ identically prepared qubits is locally equivalent to a model with only a classical Gaussian distribution and a Gaussian state of a quantum harmonic oscillator.

The term `local' refers to a shrinking neighborhood around a fixed state $\rho_{0}$. 
An essential result is that the neighborhood radius can be chosen 
arbitrarily close to $n^{-1/4}$. This allows us to use a two steps procedure by which we first localize the state within a smaller neighborhood of radius $n^{-1/2+\epsilon}$, and then use LAN to perform optimal estimation. 

}

\setlength{\parindent}{0pt}
\setlength{\parskip}{2mm}

%\tableofcontents
\section{Introduction}

State estimation is a central topic in quantum  statistical inference \cite{Holevo,Helstrom,Barndorff-Nielsen&Gill&Jupp,Hayashi.editor}.
In broad terms the problem can be formulated as follows: given a quantum system prepared in an unknown state $\rho$, one would like to reconstruct the state by performing a measurement $M$ whose random result $X$ will be used to build an estimator $\hat{\rho}(X)$ of $\rho$. The quality of the measurement-estimator pair is
given by the {\it risk}
\begin{equation}\label{eq.risk}
R_{\rho}(M, \hat{\rho} ) = \mathbb{E} \left(d(\hat{\rho} (X) , \rho)^{2}\right),
\end{equation}
where $d$ is a distance on the space of states, for instance the fidelity
distance or the trace norm, and the expectation is taken with respect to the
probability distribution $\mathbb{P}^{M}_{\rho}$ of $X$, when the measured system is in state $\rho$. Since the risk depends on the unknown state
$\rho$, one considers a global figure of merit by either averaging with respect to a prior distribution $\pi$ (Bayesian setup)
\begin{equation}\label{eq.risk.bayes}
R_{\pi}(M, \hat{\rho}) = \int \pi(d\rho) R_{\rho}(M, \hat{\rho} ),
\end{equation}
or by considering a maximum risk (pointwise or minimax setup)
\begin{equation}\label{eq.risk.max}
R_{\rm max} (M, \hat{\rho}) = \sup_{\rho}  R_{\rho}(M, \hat{\rho} ).
\end{equation}
An optimal procedure in either setup is one which achieves the minimum risk.

Typically, one measurement result does not provide enough information in
order to significantly narrow down on the true state $\rho$. Moreover, if
the measurement is ``informative'' then the state of the system after the
measurement will contain little or no  information about the initial state
\cite{bas} and one needs to repeat the preparation and measurement procedure in order
to estimate the state with the desired accuracy.

It is then natural to consider a framework in which we are given a number
$n$ of identically prepared systems and look for estimators $\hat{\rho}_{n}$ which are optimal, or become optimal in the limit of large $n$. This problem is the quantum analogue of the classical statistical problem \cite{vanderVaart} of estimating a parameter
$\theta$ from independent identically distributed random variables
$X_{1},\dots , X_{n}$ with distribution $\mathbb{P}_{\theta}$, and some of the methods developed in this paper are inspired by the classical theory.

Various state estimation problems have been investigated in the
literature and the techniques may be quite different depending on
a number of factors: the dimension of the density matrix, the number of
unknown parameters, the purity of the states, and the complexity of measurements over which one optimizes.
A short discussion on these issues can be found in section \ref{sec.estimation}.

In this paper we give an asymptotically optimal measurement strategy for qubit
states that is based on the technique of {\it local asymptotic normality} introduced 
in \cite{Guta&Kahn,Guta&Jencova}. The technique is a quantum generalisation of Le Cam's classical statistical result \cite{LeCam}, and builds on previous work of Hayashi and Matsumoto \cite{Hayashi.conference,Hayashi&Matsumoto}.
We use an adaptive two stage procedure involving
continuous time  measurements, which could in principle be implemented in
practice. The idea of adaptive estimation methods, which has a long history in classical 
statistics, was introduced in the quantum set-up by \cite{Barndorff-Nielsen&Gill}, and was subsequently used in \cite{Gill&Massar,Hayashi,Hayashi&Matsumoto2}.  The aim there is similar: one wants to first localize the state and then to perform a suitably tailored measurement which performs optimally around a given state.  A different adaptive technique was proposed independently by Nagaoka \cite{Nagaoka} and further developed in \cite{Fujiwara}.

\vspace{5mm}

\begin{figure}[h]
\begin{center}
\epsfig{file=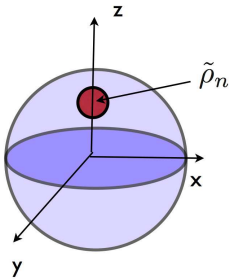, height=4cm}
\caption{After the first measurement stage the state $\rho$ lies in a small ball 
centered at $\tilde{\rho}_{n}$.}
\label{fig.small.ball}
\end{center}
\end{figure}
In the first stage,
the spin components $\sigma_x$, $\sigma_y$ and $\sigma_z$
are measured separately on a small portion
$\tilde{n} \ll n$ of the
systems, and a rough estimator $\tilde{\rho}_{n}$ is constructed.
By standard statistical arguments (see Lemma \ref{lemma.small.probability}) we deduce that with high probability, the true state
$\rho$ lies within a ball of radius slightly larger than $n^{-1/2}$, say 
$n^{-1/2 + \epsilon}$ with $\epsilon>0$, centered at
$\tilde{\rho}_{n}$. The purpose of the first stage is thus
to localize the state within a small neighborhood as illustrated in Figure 
\ref{fig.small.ball} (up to a unitary rotation) using the Bloch sphere representation 
of qubit states. 
%This information is then used in the second stage, which is a
%{\it joint} measurement on the remaining $n-\tilde{n}$ systems.
%This second measurement is implemented physically by two consecutive
%couplings, each to a bosonic field. The qubits are first coupled to the field via a spontaneous emission interaction and a continuous time heterodyne detection measurement is performed in the field. This yields information on the eigenvectors of $\rho$. Then the interaction is changed, and a continuous time homodyne detection is performed in the field. This yields information on the eigenvalues of $\rho$.

%\vspace{2cm}
%\begin{figure}[h]
%\begin{center}
%\includegraphics[height=4cm]{smallball2}
%\caption{After the first measurement stage the state $\rho$ lies in a small ball 
%centered at $\tilde{\rho}_{n}$.}
%\label{fig.small.ball}
%\end{center}
%\end{figure}
This information is then used in the second stage, which is a
{\it joint} measurement on the remaining $n-\tilde{n}$ systems.
This second measurement is implemented physically by two consecutive
couplings, each to a bosonic field. The qubits are first coupled to the field via a spontaneous emission interaction and a continuous time heterodyne detection measurement is performed in the field. This yields information on the eigenvectors of $\rho$. Then the interaction is changed, and a continuous time homodyne detection is performed in the field. This yields information on the eigenvalues of $\rho$.
 
%The largest part of this paper is devoted to proving that

\begin{figure}[h]
%\begin{wrapfigure}[h!]{r}{60mm}
\begin{center}
\epsfig{file=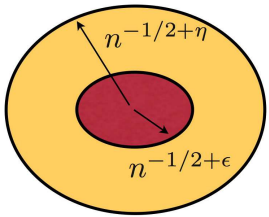, width=4cm}
\caption{The smaller domain is the localization region of the first step. The second stage estimator is optimal for all states in the bigger domain.}
\label{fig.2balls}
\end{center}
\end{figure}

We prove that the second stage of the measurement
is asymptotically optimal for all states in a ball of radius $n^{-1/2 + \eta}$  
around $\tilde{\rho}_{n}$. Here $\eta$ can be chosen to be bigger that 
$\epsilon>0$ implying that the two stage procedure as a whole is asymptotically optimal for any state as depicted in Figure \ref{fig.2balls}.

The optimality of the second stage
relies heavily on the principle of \emph{local asymptotic normality} or  LAN, see \cite{vanderVaart}, which we will briefly explain below,
and in particular on the fact that it holds in a ball of radius $n^{-1/2 + \eta}$
around $\tilde{\rho}_{n}$ rather than just $n^{-1/2}$ as it was the case in \cite{Guta&Kahn}. 

Let $\rho_{0}$ be a fixed state.
We parametrize the neighboring states as $\rho_{{\bf u}/\sqrt{n}}$,
where ${\bf u}=(u_{x},u_{y},u_{z})\in\mathbb{R}^{3}$
is a certain set of local parameters around $\rho_0$. Then LAN entails that the joint state
$\rho^{\bf u}_{n}: =\rho_{{\bf u}/\sqrt{n}}^{\otimes n} $
of $n$ identical qubits
converges for $n \rightarrow \infty$
to a Gaussian state of the form $N^{\bf u}\otimes\phi^{\bf u}$,
in a sense explained in Theorem \ref{th.qlan}. By $N^{\bf u}$ we denote a {\it classical} one-dimensional
normal distribution centered at $u_{z}$. The second term $\phi^{\bf u}$ is a Gaussian state of a harmonic
oscillator, i.e. a displaced thermal equilibrium state with
displacement proportional to $(u_{x},u_{y})$. We thus have the convergence
$$
\rho^{\bf u}_{n}\leadsto N^{\bf u}\otimes \phi^{\bf u},
$$
to a much simpler family of classical -- quantum states for which we know how
to optimally estimate the parameter {\bf u} \cite{Holevo,Yuen&Lax}.

The idea of approximating a sequence of statistical experiments by a
Gaussian one goes back to Wald \cite{Wald2}, and was subsequently developed by Le Cam \cite{LeCam} who coined the term local asymptotic normality. In quantum statistics  the first ideas in the direction of local asymptotic normality for d-dimensional 
states appeared in the Japanese paper \cite{Hayashi.japanese}, as well as \cite{Hayashi.conference} and were subsequently developed in \cite{Hayashi&Matsumoto}. In Theorem \ref{th.qlan} we strengthen these results for the case of qubits, by proving a strong version of LAN in the spirit of Le Cam's pioneering work. We then exploit this result to prove optimality of the second stage. A different approach to local asymptotic normality has been developed in \cite{Guta&Jencova} to which we refer for a more general exposition on the theory of quantum statistical models. A short discussion on the relation between the two approaches is given in the remark following Theorem \ref{th.qlan}.

%Another approach to local asymptotic normality has been investigated in \cite{Guta&Jencova}, where instead of  

\begin{figure}[h!]
\begin{center}
\epsfig{file=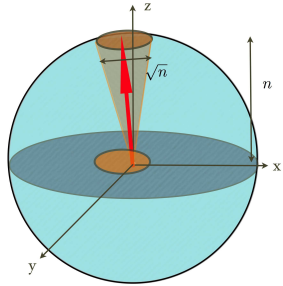, width=4cm}
\caption{Total spin representation of the state of $n\gg 1$  spins: the quantum fluctuations of the $x$ and 
$y$ spin directions coincide with those of a coherent state of a harmonic oscillator.}
\label{Fig.rotated.spins}
\end{center}
\end{figure}

From the physics perspective, our results put on a more rigorous basis the treatment 
of collective states of many identical spins, the keyword here being {\it coherent spin states} \cite{Holtz}. Indeed, it has been known since Dyson \cite{Dyson1} that $n$ spin-$\frac{1}{2}$ particles prepared in the spin up state $|\!\uparrow\rangle^{\otimes n}$ behave asymptotically as the ground state of a quantum oscillator, when considering the fluctuations of properly normalized total spin components in the directions orthogonal to $z$. We extend this to spin directions making an ``angle'' of order $n^{-1/2 + \eta}$ with the $z$ axis, as illustrated in Figure \ref{Fig.rotated.spins}, as well as to mixed states. We believe that a similar approach can be followed in the case of spin squeezed states and continuous time measurements with 
feedback control \cite{GSM}.

 In Theorem \ref{th.unitary.evolution} we prove a dynamical version of LAN.
 The trajectory in time of the joint state of the qubits together with the field  converges for large  $n$ to the corresponding trajectory of the joint state of the oscillator and field.
 In other words, time evolution preserves local asymptotic normality.
 This insures that for large $n$  the state of the qubits ``leaks'' into a Gaussian state of the field, providing a concrete implementation of the convergence to the limit Gaussian experiment.

The punch line of the paper is Theorem \ref{automobiel} which says that the
estimator $\hat\rho_{n}$ is optimal in local minimax sense, which is the
modern statistical formulation of optimality in the frequentist setup \cite{vanderVaart}.
Also, its asymptotic risk is calculated explicitly.

%The asymptotic splitting into
%a classical estimation problem for eigevalues and a quantum one for the
%eigenbasis has been also noticed in \cite{Bagan&Gill} and in
%\cite{Hayashi&Matsumoto}, the latter coming pretty close to our
%formulation of local asymptotic normality.

The paper is structured as follows: in section \ref{sec.estimation},
we show that the first stage of the measurement sufficiently
localizes the state. In section \ref{sec.lan}, we prove
that LAN holds with radius of validity $n^{-1/2 + \eta}$, and we bound
its rate of convergence.
sections \ref{sec.timevolf} and \ref{finmes} are concerned with the second stage of the measurement, i.e. with the coupling to the bosonic field and the continuous time
field-measurements. Finally, in section \ref{sec.endresult},
asymptotic optimality of the estimation scheme is proven.

The technical details of the proofs are relegated to the appendices in order to give the reader a more direct access  to the ideas and results.

\section{State estimation}\label{sec.estimation}

In this section we introduce the reader to a few general aspects of quantum
state estimation after which we concentrate on the qubit case.

State estimation is a generic name for a variety of results which may be classified
according to the dimension of the parameter space, the kind or family of states to be estimated and the preferred estimation method. For an introduction to quantum statistical 
inference we refer to the books by Helstrom \cite{Helstrom} and Holevo \cite{Holevo} and the more recent review paper \cite{Barndorff-Nielsen&Gill&Jupp}. The collection \cite{Hayashi.editor} is a good reference on quantum statistical problems, with many important contributions by the Japanese school.

For the purpose of this paper, any quantum state representing a particular preparation of a quantum system, is described by a density matrix (positive selfadjoint operator of trace one) on the Hilbert space $\mathcal{H}$ associated to the system. The algebra of observables is $\mathcal{B}(\mathcal{H})$, and the expectation of an observable 
$a\in\mathcal{B}(\mathcal{H})$ with respect to the state $\rho$ is 
$\mathrm{Tr}(\rho a)$. A measurement $M$ with outcomes in a measure space 
$(\mathcal{X},\Sigma)$ is completely determined by a $\sigma$-additive collection of positive selfadjoint operators $M(A)$ on $\mathcal{H}$, where $A$ is an event in $\Sigma$. This collection is called a positive operator valued measure. The distribution of the results $X$ when the system is in state $\rho$ is given 
by $P_{\rho}(A) = \mathrm{Tr}(\rho M(A))$.  

We are given $n$ systems identically prepared in state $\rho$ and we are allowed to perform a measurement $M_{n}$ whose outcome is the estimator $\hat{\rho}_{n}$ as discussed in the Introduction.
 
The dimension of the density matrix may be finite, such as in the case of qubits or d-levels atoms, or infinite as in the case of the state of a monochromatic beam of
light. In the finite or parametric case one expects that the risk converges to zero as
$n^{-1}$ and the optimal measurement-estimator sequence $(M_{n},\hat{\rho}_{n})$ achieves the {\it best constant} in front of the $n^{-1}$ factor.
In the non-parametric case the rates of convergence are in general slower that
$n^{-1}$ because one has to simultaneously estimate an infinite number of matrix elements, each with rate $n^{-1}$.
An important example of such an estimation technique is that of quantum homodyne tomography in quantum optics \cite{Vogel&Risken}. This allows the estimation with arbitrary precision \cite{D'Ariano.2, D'Ariano.3,Leonhardt.Munroe} of the whole density matrix  of a monochromatic beam of light by repeatedly measuring a sufficiently large number of identically prepared beams \cite{Smithey,Breitenbach&Schiller&Mlynek,Zavatta}. In \cite{Artiles&Guta&Gill,Butucea&Guta&Artiles} it is shown how to formulate the
problem of estimating infinite dimensional states without the need for choosing a cut-off in the dimension of the density matrix, and how to construct optimal minimax estimators of the Wigner function for a class of ``smooth'' states.

If we have some prior knowledge about the preparation procedure, we may encode
this by parametrizing the possible states as $\rho=\rho_{\theta}$ with
$\theta\in\Theta$ some unknown parameter. The problem is then to
estimate
$\theta$ optimally with respect to a distance function on $\Theta$.
%For example if the state is pure or belongs to a one parameter family of states,
%the optimal measurement turns out to be simpler.

Indeed, one of the main problems in the finite dimensional case is to find optimal estimation procedures for a given family of states. It is known that if the state $\rho$ is pure or belongs to a one parameter family, then separate measurements achieve
the optimal rate of the class of joint measurements \cite{Matsumoto}.
However for multi-dimensional families of mixed states this is no longer the case and joint measurements perform strictly better than separate ones \cite{Gill&Massar}.

In the Bayesian setup, one optimizes $R_{\pi}(M_{n},\hat{\rho}_{n})$ for some prior distribution $\pi$. We refer to \cite{Jones,Massar&Popescu,Latorre&Pascual&Tarrach,Fisher&Kienle&Freyberger,
Wunderlich,Bagan&Baig&Tapia,Narnhofer,Bagan&Monras&Tapia} for the pure state case, and to \cite{Cirac,Vidal,Mack, Keyl&Werner,Bagan&Baig&Tapia&Rodriguez,Sommers,Bagan&Gill} for the
mixed state case. The methods used here are based on group theory and can be applied only to invariant prior distributions and certain distance functions.
In particular, the optimal covariant measurement in the case of completely unknown
qubit states was found in \cite{Bagan&Gill,Hayashi&Matsumoto} but it has the drawback that it does not give any clue as to how it can be
implemented in a real experiment.

In the pointwise approach \cite{Hayashi,Hayashi&Matsumoto2,Gill&Massar,Barndorff-Nielsen&Gill, Fujiwara&Nagaoka,Matsumoto,Barndorff-Nielsen&Gill&Jupp,Hayashi&Matsumoto} one tries to minimize the risk for {\it each} unknown state $\rho$. As the optimal measurement-estimator pair cannot depend on 
the state itself, one optimizes the maximum risk $R_{\rm max}(M_{n},\hat{\rho}_{n})$, (see \eqref{eq.risk.max}), or a local version of this which will be defined shortly. The advantage of the pointwise approach is that it can be applied to arbitrary families of states and a large class of loss functions provided that they are locally quadratic in the chosen parameters.  The underlying philosophy is that as the number $n$ of states is sufficiently large, the problem ceases to be global and becomes a local one as the error in estimating the state parameters is of the order $n^{-1/2}$. 
%The Japanese school has brought important contributions in this area and many papers, some of which were previously published in Japanese, are now available in \cite{Hayashi.editor}.

The Bayesian and pointwise approaches can be compared
\cite{Gillunpub}, and in fact for large $n$ the prior distribution $\pi$ of the Bayesian approach becomes increasingly irrelevant and the optimal Bayesian estimator becomes asymptotically optimal in the minimax sense and vice versa.

\subsection{Qubit state estimation: the localization principle}

Let us now pass to the quantum statistical model which will be the object of our investigations. Let $\rho \in M_{2}(\mathbb{C})$ be an arbitrary density matrix describing the state of a qubit. Given $n$ identically prepared qubits with joint state $\rho^{\otimes n}$, we would like to optimally estimate $\rho$ based on the result of a properly
chosen joint measurement $M_{n}$. For simplicity of the exposition we assume that
the outcome of the measurement is an estimator $\hat \rho_{n}\in M_{2}(\mathbb{C})$. In practice however, the result $X$ may belong to a complicated measure space (in our case the space of continuous time paths) and the estimator is a function of the ``raw'' data $\hat{\rho}_{n} := \hat{\rho}_{n}(X)$. The quality of the estimator at the state $\rho$ is quantified by the risk
$$
R_{\rho}(M_{n}, \hat{\rho}_{n}) := \mathbb{E}_{\rho} (d (\rho, \hat{\rho}_{n})^{2}),
$$
where $d$ is a distance between states. The above expectation is taken with respect to
the distribution $P_{\rho}(dx):=\mathrm{Tr}(\rho M(dx))$ of the measurement results, where $M(dx)$ represents the associated positive operator valued measure of the measurement $M$. 
In our exposition $d$ will be the trace norm
$$
\| \rho_{1} - \rho_{2}\|_{1} :=  \mathrm{Tr}(|\rho_{1} - \rho_{2}|),
$$
but similar results can be obtained using the fidelity distance. The aim is to find a sequence of measurements and estimators $(M_{n}, \hat{\rho}_{n})$ which is asymptotically optimal in the {\it local minimax} sense: for any given $\rho_{0}$
$$
\limsup_{n\to \infty}\sup_{\|\rho - \rho_0\|_{1} \leq n^{-1/2 + \epsilon}}  n R_{\rho}(M_{n}, \hat{\rho}_{n}) \leq
\limsup_{n\to \infty}\sup_{\|\rho - \rho_0\|_{1} \leq n^{-1/2 + \epsilon}}  n R_{\rho}(N_{n}, \check\rho_{n}),
$$
for any other sequence of measurement-estimator pairs $(N_{n},\check{\rho}_{n})$. The factor $n$ is inserted because typically $R_{\rho}(M_{n}, \hat{\rho}_{n})$ is of the order $1/n$ and the optimization is about obtaining the smallest constant factor possible.
The inequality says that one cannot find an estimator which performs better that
$\hat\rho_{n}$ over a ball of size $n^{-1/2 + \epsilon}$ centered at $\rho_{0}$,
even if one has the knowledge that  the state $\rho$ belongs to that ball! 

Here, and elsewhere in the paper $\epsilon$ will appear in different contexts, as a generic strictly positive number and will be chosen to be sufficiently small for each specific use. At places where such notation may be confusing we will use additional symbols to denote small constants.

As set forth in the Introduction,
our measurement procedure consists of two steps. The first one is to perform
separate measurements of $\sigma_x$, $\sigma_y$ and $\sigma_z$
%, identical measurements $\tilde M$
on a fraction $\tilde{n}=\tilde{n}(n)$ of the systems.
In this way we
obtain a rough estimate $\tilde \rho_{n}$ of the true state $\rho$
which lies in a local neighborhood around $\rho$ with
high probability.
The second step uses the information obtained in the first step
to perform a measurement which is optimal precisely for the states
in this local neighborhood.
The second step ensures optimality and requires more sophisticated techniques
inspired by the theory of local asymptotic normality for qubit states \cite{Guta&Kahn}.
We begin by showing that the first step amounts to the fact that, without loss of
generality, we may assume that the unknown state is in a local neighborhood of a known state. This may serve also as an a posteriori justification of the definition of local
minimax  optimality.
\begin{lemma}\label{lemma.small.probability}
Let $M_{i}$ denote the measurement of the $\sigma_{i}$ spin component of a
qubit with $i=x,y,z$. We perform each of the measurements $M_{i}$ separately
on $\tilde{n}/3$ identically prepared qubits and define
$$
\tilde\rho_{n} = \frac{1}{2}(\mathbf{1} + \tilde{\mathbf{r}} \sigma),\qquad {\rm if~}\quad |\tilde{r}|\leq 1,
$$
where $\tilde{\mathbf{r}} =(\tilde r_{x}, \tilde r_{y}, \tilde r_{z})$ is the vector average
of the measured components. If $| \tilde{r} |>1$ then we define $\tilde\rho_{n}$ as the state which has the smallest trace distance to the right hand side expression. 
Then for all
%$a \in \mathbb{R}$,
$\epsilon \in [0,2]$,
we have
$$
\mathbb{P} \left(\| \tilde\rho_{n} -\rho \|_{1}^{2} >
%3 \frac{n^{a}}{n} \right) \leq
3 n^{2 \epsilon-1} \right) \leq
%6 \exp(-\half \tilde{n} n^{a - 1} ), \qquad \forall \rho.
6 \exp(-\half \tilde{n} n^{2 \epsilon - 1} ), \qquad \forall \rho.
$$
Furthermore, for any
%$b\in (0,1)$ such that $a+b>1$, if $\tilde{n}=n^{b}$
$0< \kappa < \epsilon/2$, if $\tilde{n}=n^{1-\kappa}$, the contribution to the risk 
$\mathbb{E} (\| \tilde{\rho}_{n}-\rho\|_{1}^{2})$ brought by the event 
$E= [\, \| \tilde\rho_{n} -\rho \|_{1} > \sqrt{3} n^{-1/2 +\epsilon}\,] $ satisfies
$$
\mathbb{E} \left(\, \| \tilde{\rho}_{n} - \rho \|_{1}^{2}\, \chi_{E}\, \right) \leq 24 \exp(-\half n^{2 \epsilon - \kappa}) = o(1).
$$
\end{lemma}

\noindent{\it Proof.}
For each spin component $\sigma_{i}$ we obtain i.i.d coin tosses $X_{i}$ with distribution
$\mathbb P(X_{i} = \pm 1) =(1\pm r_{i})/2$ and average $r_{i}$.

Hoeffding's inequality \cite{vanderVaart&Wellner} then states that for all $c > 0$, we have
$\mathbb{P}( |X_i - \tilde{X} |^2 > c ) \leq 2 \exp(- \half \tilde{n}c ) $.
By using this inequality  three times with
%$c = n^{a-1}$,
$c = n^{2 \epsilon -1}$,
once for each component, we get
$$
\mathbb{P} \left( \sum_{1}^{3} | \tilde r_{i} -r_{i}|^{2} > 3
n^{2 \epsilon-1}  \right) \leq
%6 \exp(-\half \tilde{n} n^{a - 1} )
6 \exp(-\half \tilde{n} n^{2 \epsilon - 1} )
  \qquad \forall \rho,
$$
which implies the statement for the norm distance since 
$\|\tilde{\rho}_{n} -\rho \|_{1}^{2} =\sum_{i} |\tilde{r}_{i} -r_{i}|^{2}$.
The bound on conditional risk follows from the previous bound and the fact that
$\| \rho- \tilde{\rho}_{n} \|_{1}^{2} \leq 4$.

\qed

In the second step of the measurement procedure we rotate the
remaining $n-\tilde{n}$ qubits such that after rotation the vector $\tilde r$ is parallel to the $z$-axis. Afterwards,
we couple the systems to the field and perform certain measurements in the field which will
determine the final estimator $\hat{\rho}_{n}$.
The details of this second step are given in sections \ref{sec.timevolf} and
\ref{finmes}, but at this moment we can already prove that the effect of errors in the the first stage of the measurement is
asymptotically negligible compared to the risk of the second estimator. Indeed by Lemma
\ref{lemma.small.probability} we get that if
%$\tilde{n}= n^{b}$,
$\tilde{n}= n^{1 - \kappa}$,
then the probability that the first stage gives a
``wrong'' estimator (one which lies outside the local neighborhood of the  true state)  is of the order
$\exp(-\half n^{2 \epsilon - \kappa})$ and so is the risk contribution. As the typical risk of
estimation is of the order $1/n$, we see that the first step is practically ``always''
placing the estimator in a neighborhood of order
%$n^{(a - 1)/2}$
$n^{- 1/2 + \epsilon}$
of the true state $\rho$, as shown in Figure \ref{fig.2balls}.
In the next section we will show that for such neighborhoods, the state of the remaining
$n-\tilde{n}$ systems behaves asymptotically as a Gaussian state. This will allow us to devise an optimal
measurement scheme for qubits based on the optimal measurement for Gaussian states.

\section{Local asymptotic normality}\label{sec.lan}

The optimality of the second stage of the measurement
relies on the concept of local asymptotic normality \cite{vanderVaart,Guta&Kahn}.
After a short introduction, we will prove that LAN holds for the qubit
case, with radius of validity $n^{-1/2 + \eta}$ for all $\eta \in [0,1/4)$.
We will also show that its rate of convergence is
$O(n^{-1/4 + \eta + \epsilon})$ for arbitrarily
small $\epsilon$.

\subsection{Introduction to LAN and some definitions}

Let $\rho_{0}$ be a fixed state, %(this is the first stage estimator $\tilde{\rho}_{n}$)
%which by the rotational symmetry may be chosen to be diagonal with respect
%to the $\sigma_{z}$ basis.
which by rotational symmetry can be chosen of the form
\begin{equation}
\label{rho0}
\rho_{0} =\left(
\begin{array}{cc}
\mu & 0\\
0 & 1-\mu
\end{array}
\right),
\end{equation}
for a given $\frac{1}{2}<\mu< 1$.
We parametrize the neighboring states as $\rho_{{\bf u}/\sqrt{n}}$ where
${\bf u}=(u_{x},u_{y},u_{z})\in\mathbb{R}^{3}$ such that the first two
components  account for unitary rotations around $\rho_{0}$, while the
third one describes the change in eigenvalues
%for some parametrization (see equation \eqref{eq.family}).
  \begin{equation}\label{eq.family}
\rho_{\bf v} := U\left({\bf v}\right)
\left(
\begin{array}{cc}
\mu + v_{z} & 0\\
0 & 1-\mu -v_{z}
\end{array}
\right)
U\left({\bf v} \right)^{*}  , %\quad{\bf u}=(u_{x},u_{y},u_{z}) \in \mathbb{R}^{3},
\end{equation}
with unitary
$U({\bf v}):= \exp(i(v_{x}\sigma_{x}  + v_{y}\sigma_{y}))$.
The ``local parameter'' ${\bf u}$ should be thought of, as having a
bounded range in $\mathbb{R}^{3}$ or may even ``grow slowly''
as $\|{\bf u}\|\leq n^{\eta}$.

Then, for large $n$, the joint state
$\rho^{\bf u}_{n}: =\rho_{{\bf u}/\sqrt{n}}^{\otimes n} $ of $n$ identical qubits
approaches a Gaussian state of the form $N^{\bf u}\otimes\phi^{\bf u}$ with
the parameter ${\bf u}$ appearing solely in the average of the two Gaussians.
By $N^{\bf u}$ we denote a {\it classical} one-dimensional
normal distribution centered at $u_{z}$ which relays
information about the eigenvalues of $\rho_{{\bf u}/\sqrt{n}}$.
The second term $\phi^{\bf u}$ is a Gaussian state of a harmonic
oscillator which is a displaced thermal equilibrium state with
displacement proportional to $(u_{x},u_{y})$.
It contains information on the eigenvectors of $\rho_{{\bf u}/\sqrt{n}}$.
We thus have the convergence
$$
\rho^{\bf u}_{n}\leadsto N^{\bf u}\otimes \phi^{\bf u},
$$
to a much simpler family of classical - quantum states for which we know how
to optimally estimate the parameter {\bf u}. The asymptotic splitting into
a classical estimation problem for eigenvalues and a quantum one for the
eigenbasis has been also noticed in \cite{Bagan&Gill} and in
\cite{Hayashi&Matsumoto}, the latter coming pretty close to our
formulation of local asymptotic normality.

The precise meaning of the convergence is given in Theorem \ref{th.qlan}
below.
In short, there exist quantum channels $T_{n}$ which map the states
$\rho_{{\bf u}/\sqrt{n}}^{\otimes n}$
into $N^{\bf u}\otimes\phi^{\bf u}$ with vanishing error in trace norm
distance, and uniformly over the local parameters ${\bf u}$. From the
statistical point of view the convergence implies
that a statistical decision problem concerning the model $\rho^{\bf u}_{n}$
can be mapped into a similar problem for the model
$N^{\bf u}\otimes\phi^{\bf u}$ such that the optimal solution for the
latter  can be translated into an asymptotically optimal solution
for the former. In our case the problem of estimating the state
$\rho$ turns into that of estimating the local parameter ${\bf u}$ around
the first stage estimator
$\tilde{\rho}_{n}$ playing the role of $\rho_{0}$. For the family of
displaced Gaussian states it is well known that the optimal estimation
of the displacement is achieved by the heterodyne detection
\cite{Holevo,Yuen&Lax}, while for the classical part it sufficient to
take the observation as best estimator. Hence the second step will
give an optimal estimator $\hat{\bf u}$ of ${\bf u}$ and an optimal
estimator of the initial state
$\hat{\rho}_{n}:= \rho_{\hat{\bf u}/\sqrt{n}}$. The precise result
is formulated in Theorem \ref{automobiel}
%\ref{th.local.minimax.est}.

%Another approach to local asymptotic normality has been investigated in 
%\cite{Guta&Jencova}. 

%%%%%%%%%%%%%%%%%%%%%%%%%%%%%%%%%%%%%%%%%%%%%%%%%%%%%%%%%%%%%%%%%%%%

\subsection{Convergence to the Gaussian model}

We describe the state $N^{\bf u} \otimes \phi^{\bf u}$ in more detail.
$N^{\bf u}$ is simply the classical Gaussian distribution
\begin{equation} \label{gausje}
N^{\bf u}:=N(u_{z}, \mu(1-\mu)),
\end{equation}
with mean $u_z$ and variance $\mu(1-\mu)$.

The state $\phi^{\bf u}$ is a density matrix on
$\mathcal{H} = \mathcal{F}(\mathbb{C})$,
the representation space of the harmonic oscillator.
In general, for any Hilbert space $\mathfrak{h}$,
the {\it Fock space} over $\mathfrak{h}$ is defined as
\begin{equation}\label{eq.Fock}
\mathcal{F}(\mathfrak{h}) := \bigoplus_{n=0}^{\infty}
\mathfrak{h} \otimes_{s}\dots
\otimes_{s} \mathfrak{h},
\end{equation}
with $\otimes_{s}$ denoting the symmetric tensor product.
Thus $\mathcal{F}(\mathbb{C})$ is the simplest example of a Fock space.
Let
\begin{equation}
\label{phi0}
\phi := (1-p) \sum_{k=0} p^{k} |k\rangle \langle k|,
\end{equation}
be a thermal equilibrium state with $|k\rangle$ denoting the $k$-th energy level of
the oscillator and $p=\frac{1-\mu}{\mu}<1$.
For every $\alpha\in\mathbb{C}$ define the
displaced thermal state
\begin{equation*}
\label{phiu}
\phi(\alpha):= D( \alpha)  \,\phi\, D(-\alpha),
\end{equation*}
where $D(\alpha): = \exp( \alpha a^{*} - \bar \alpha a)$
is the displacement operator, mapping
the vacuum vector $| 0\rangle$ to the coherent vector
$$
| \alpha \rangle = \exp( - \alpha^2 \!/ 2)\sum_{k=0}^{\infty}
\frac{\alpha^k}{\sqrt{k!}} |k\rangle. 
$$
Here $a^{*}$ and $a$ are the creation and annihilation operators on
$\mathcal{F}(\mathbb{C})$,
satisfying $[a, a^{*}] = \mathbf{1}$.
The family $\phi^{\bf u}$ of states in which we are interested
%depends on the first two coordinated of ${\bf u}$ and
is given by
\begin{equation}\label{eq.displacedthermal}
\phi^{\bf u} := \phi(\sqrt{2\mu -1} \alpha_{\bf u}) ,\qquad {\bf u} \in\mathbb{R}^{3},
\end{equation}
with $\alpha_{\bf u}:= -u_{y}+ iu_{x}$. Note that $\phi^{\bf u}$ does not depend on $u_z$.

We claim that the ``statistical information'' contained in the joint
state of $n$ qubits
\begin{equation}\label{eq.family.n}
\rho_n^{\bf u} :=  \rho_{{\bf u}/\sqrt{n}}^{\otimes n},
\end{equation}
is asymptotically identical to that contained in the couple
$(N^{\bf{u}} , \phi^{\bf u})$. More precisely:

\begin{theorem}\label{th.qlan}
Let $\rho_n^{\bf u}$ be the family of states \eqref{eq.family} on the
Hilbert space
$\left( \mathbb{C}^{2} \right)^{\otimes n}$, let $N^{\mathbf{u}}$ be the family
\eqref{gausje} of Gaussian distributions, and let $\phi^{\bf u}$ be the family
\eqref{eq.displacedthermal} of displaced thermal equilibrium states of a quantum
oscillator. Then for each $n$ there exist quantum channels (trace preserving CP maps)
\begin{equation*}
\begin{split}
T_{n} :
%M_{2}^{\otimes n}
\mathcal{T}((\mathbb{C}^{2})^{\otimes n})
\to
           L^{1}(\mathbb{R})\otimes \mathcal{T}(\mathcal{F}(\mathbb{C})), \\
S_{n} :  L^{1}(\mathbb{R})\otimes\mathcal{T}(\mathcal{F}(\mathbb{C})) \to
\mathcal{T}((\mathbb{C}^{2})^{\otimes n})
%M_{2}^{\otimes n},
\end{split}
\end{equation*}
with $\mathcal{T}(\mathcal{H})$ the trace-class operators on $\mathcal{H}$,
such
that, for any $0 \leq \eta<1/4$ and any $\epsilon> 0$,
\begin{eqnarray}\label{eq.channel.conv.}
&&
\sup_{\|{\bf u}\| \leq n^{\eta}} \,
\|  N^{\bf u}\otimes \phi^{\bf u} - T_{n} \left(  \rho^{\bf u}_{n}\right)
\|_{1} = O(n^{-1/4+\eta + \epsilon} ), \\
&&
\sup_{\|{\bf u}\| \leq n^{\eta}} \,
\| \rho^{\bf u}_n - S_{n} \left(  N^{\bf u}\otimes \phi^{\bf u}\right)  \|_{1}
= O(n^{-1/4+ \eta + \epsilon }). \label{eq.channel.conv.inverse}
\end{eqnarray}
Moreover, for each $\epsilon_2 > 0$ there exists a function $f(n)$ of order
$O(n^{-1/4+\eta + \epsilon} )$ such that the above convergence rates are
bounded by $f(n)$, with $f$ independent of $\rho^{\mathbf{0}}$ as long as
$|\half - \mu| > \epsilon_2$.
\end{theorem}

\noindent
{\bf Remark.} Note that the equations \eqref{eq.channel.conv.} and 
\eqref{eq.channel.conv.inverse} imply that the expressions on the 
left side converge to zero as $n\to\infty$. 
Following the classical terminology of Le Cam \cite{LeCam}, we will call this type of result 
{\it strong convergence} of quantum statistical models (experiments). 
Another local asymptotic normality result has been derived in \cite{Guta&Jencova} 
based on a different concept of convergence, which is  an extension of the 
{\it weak convergence} of classical (commutative) statistical experiments. 
In the classical set-up it is known that strong convergence implies weak convergence for arbitrary statistical models, and the two are equivalent for statistical models consisting of a finite number of distributions. A similar relation is conjectured to hold in the quantum set-up, but for the moment this has been shown only under additional assumptions 
\cite{Guta&Jencova}. 

These two approaches to local asymptotic normality in quantum statistics are based on completely different methods and the results are complementary in the sense that the weak convergence of \cite{Guta&Jencova} holds for the larger class of finite dimensional states while the strong convergence has more direct consequences as it is shown 
in this paper for the case of qubits. Both results are part of a larger effort to develop a general theory of local asymptotic normality in quantum statistics. Several extensions are in order: from qubits to arbitrary finite dimensional systems (strong convergence), from finite dimensional to continuous variables systems, from identical system to correlated ones, and asymptotic normality in continuous time dynamical set-up.

Finally, let us note that the development of a general theory of convergence of 
quantum statistical models will set a framework for dealing with other important 
statistical decision problems such as quantum cloning \cite{Werner} and quantum amplification \cite{Caves}, which do not necessarily involve measurements.

\noindent
{\bf Remark.} The construction of the channels $T_{n}, S_{n}$ in the case of
fixed eigenvalues $(u_{z} =0)$ is given in Theorem 1.1 of \cite{Guta&Kahn}. It
is also shown that a similar result holds uniformly over
$\| {\bf u}\|<C $ for any fixed finite constant $C$.
In \cite{Guta&Jencova}, it is shown that such maps also exist in the general
case, with unknown eigenvalues. A classical component then appears in the limit statistical experiment. In the above result we extend the domain of validity of these Theorems from ``local'' parameters $\|{\bf u}\|<C $ to ``slowly growing'' local
neighborhoods $\|{\bf u}\| \leq n^{\eta}$ with $\eta<1/4$.
Although this may be seen as merely a
technical improvement, it is in fact essential in order to insure
that the result of the first step of the estimation will, with high probability, fall
inside a neighborhood $\|{\bf u}\| \leq n^{\eta}$ for which local
asymptotic normality still holds (see Figure \ref{fig.2balls}).\\

%\noindent
%{\bf Remark.} The totally mixed state with $\mu=1/2$ is a singular point in the parameter space, and Theorem \ref{th.qlan} does not apply in this case. This does not mean that local asymptotic normality does not hold, but that we need to work with a different set of coordinates. This issue becomes more important for higher dimensional systems 
%where the eigenvalues may exhibit more complicated multiplicities, and will be dealt with in that context.  

\noindent{\it Proof.} Following \cite{Guta&Kahn} we will first indicate how the channels
$T_{n}$ are constructed. The technical details of the proof can be found in
Appendix \ref{sec.proof.qlan}.

The space $\left( \mathbb{C}^{2}\right)^{\otimes n}$ carries two unitary representations.
The representation $\pi_{n}$ of $SU(2)$ is given by
$\pi_{n}(u) = u^{\otimes n}$ for any $u\in SU(2)$,
and the representation $\tilde{\pi}_{n}$ of the symmetric group $S(n)$ is given
by the permutation of factors
$$
\tilde{\pi}_{n} (\tau) : v_{1} \otimes \dots \otimes  v_{n }
\mapsto v_{\tau^{-1}(1)} \otimes \dots \otimes v_{\tau^{-1}(n)}, \qquad\tau\in S(n).
$$
As $[\pi_{n} (u) , \tilde{\pi}_{n} (\tau)] =0$ for all $u\in SU(2), \tau \in S(n)$,
we have the decomposition
\begin{equation}\label{eq.decomposition}
\left( \mathbb{C}^{2}\right)^{\otimes n} =
\bigoplus_{j=0, 1/2}^{n/2} \mathcal{H}_{j} \otimes \mathcal{H}^{j}_{n}.
\end{equation}
The direct sum runs over all positive (half)-integers $j$ up to $n/2$.
For each fixed $j$,
$\mathcal{H}_{j} \cong \mathbb{C}^{2j+1}$ is an irreducible representation $U_{j}$ of
$SU(2)$ with total angular momentum \mbox{$J^{2} = j(j+1)$}, and
$\mathcal{H}^{j}_{n}\cong \mathbb{C}^{n_{j} }$ is the irreducible representation
of the symmetric group $S(n)$ with $n_{j}=\binom{n}{n/2 -j} - \binom{n}{n/2-j-1} $.
The density matrix $\rho_n^{\bf u}$ is invariant under
permutations and can be decomposed as a mixture of ``block'' density matrices
\begin{equation}
\label{blocks}
\rho_n^{\bf u} = \bigoplus_{j=0, 1/2}^{n/2}  p_{n,{\bf u}}(j) \,\rho^{\bf u}_{j,n} \otimes
\frac{\mathbf{1}}{n_{j}} \,.
\end{equation}
The probability distribution $p_{n,{\bf u}}(j)$ is given by \cite{Bagan&Gill}:
\begin{equation}
\label{pnj}
p_{n,{\bf u}}(j) := \frac{n_{j}}{2\mu_{\bf u} -1} \left(1-\mu_{\bf u}\right)^{\frac{n}{2} - j}
\mu_{\bf u}^{\frac{n}{2} + j +1}
\left(1-p_{\bf u}^{2j +1}\right),
\end{equation}
with $\mu_{\bf u}:= \mu + u_{z} /\sqrt{n}$, $p_{\bf u}:= \frac{1-\mu_{\bf u}}{\mu_{\bf u}} $.
We can rewrite $p_{n,{\bf u}}(j)$ as
\begin{equation}\label{eq.pnj}
p_{n,{\bf u}}(j) := B_{n,\mu_{\bf u}} (n/2+j)\times K(j,n,\mu,{\bf u}),
\end{equation}
where
$$
B_{n,\nu} (k) :=  \binom{n}{k}  \nu^{k} \left(1-\nu \right)^{n-k} , \qquad k= 0, \dots, n
$$
is a binomial distribution, and the factor $ K(j,n,\mu,{\bf u})$ is given by
$$
K(j,n,\mu,{\bf u}):=
\left(1 - p_{\bf u}^{2j +1}\right)
\frac{n+ (2(j-j_{n} -\sqrt{n}u_{z}) +1)/(2\mu_{\bf u} -1) }{n + (j-j_{n} -\sqrt{n}u_{z}+1)/\mu_{\bf u}}, \quad
j_{n} := n(\mu-1/2).
$$
Now $K(j,n,\mu,{\bf u}) = 1 + O(n^{-1/2+\epsilon})$
on
the relevant values of $j$, i.e. the ones in an interval of order $n^{1/2+\epsilon}$ around
$j_{n}$, as long as $\mu_{\bf u}$ is bounded away from $1/2$, which is
automatically so for big $n$. As $B_{n,\mu_{\bf u}} (k) $ is the distribution of a sum of
i.i.d. Bernoulli random variables, we can now use standard local asymptotic normality
results \cite{vanderVaart} to conclude that if $j$ is distributed according to $p_{n,{\bf u}}$, then
the centered and rescaled variable
$$
g_{n} := \frac{j}{\sqrt n} - \sqrt{n}(\mu - 1/2),
$$
converges in distribution to a normal $N^{\bf u}$, after an additional randomization has been performed.
The latter is necessary in order to ``smooth''  the discrete distribution into a distribution which is continuous with respect to the Lebesgue measure, and will convergence to the Gaussian distribution in total variation norm.

The measurement ``which block'', corresponding to the decomposition
\eqref{blocks}, provides us with a result $j$ and a posterior state $\rho^{\bf u}_{j,n}$.
The function $g_{n}= g_{n}(j)$ (with an additional randomization) is the classical part of the channel $T_{n}$.
The randomization consists of ''smoothening'' with a Gaussian kernel of mean
$g_{n}(j)$ and variance $1/(2\sqrt{n})$, i.e. with
$\tau_{n,j} :=  (n^{1/4}/\sqrt{\pi})\exp\left(-\sqrt{n} (x-g_n(j))^2\right)$.

  Note that this measurement is not disturbing the state
$\rho_n^{\bf u}$ in the sense that the average state after the measurement is the same as before. 

The quantum part of $T_{n}$ is the same as in \cite{Guta&Kahn} and consists of  embedding each block state $\rho^{\bf u}_{j,n}$ into the state space
of the oscillator by means of an isometry $V_{j} : \mathcal{H}_{j} \to\mathcal{F}(\mathbb{C})$,
$$
V_{j}: |j,m \rangle \mapsto |j-m \rangle,
$$
where $\{ |j,m \rangle : m=-j,\dots, j\}$ is the eigenbasis of the total spin component
$L_{z}:= \sum_{i}\sigma^{(i)}_{z}$, cf. equation (5.1) of \cite{Guta&Kahn}.
Then the action of the
channel $T_{n}$ is
\[
T_n: \bigoplus_{j} p_{n,{\bf u}}(j) \rho_{j,n}^{\bf u} \otimes \frac{{\bf 1}}{
n_j} \mapsto \sum_{j} p_{n,{\bf u}}(j) \, \tau_{n,j} \otimes
V_{j}\rho_{j,n}^{\bf u} V^*_{j}\,.
\]

The inverse channel $S_{n}$ performs the inverse operation with respect to
$T_{n}$. First the oscillator state is ``cut-off'' to the dimension of an irreducible representation
and then a block obtained in this way is placed into the decomposition
\eqref{eq.decomposition} (with an additional normalization from the remaining
infinite dimensional block which is negligible for the states in which we are interested).
%The detailed construction of $S_{n}$ can be found in \cite{Guta&Kahn}.

The rest of the proof is given in Appendix \ref{sec.proof.qlan}.

\qed

%
%\begin{center}
%$
%\CD
%\CS(M_2^{\otimes n}) @>>> \CS(M_2^{\otimes n} \otimes \F) @>U>>
%\CS(M_2^{\otimes n} \otimes \F) @>>> \CS({\cal C}) \\
%@VVV  @VVV  @VVV  @|  \\
%\CS(\F(\CC)) @>>> \CS(\F(\CC) \otimes \F) @>U>>
%\CS(\F(\CC) \otimes \F) @>>> \CS({\cal C})
%\endCD
%$
%\end{center}

\section{Time evolution of the interacting system}    \label{sec.timevolf}

In the previous section, we have investigated the asymptotic
equivalence between the states $\rho_{n}^{\bf u}$ and
$N^{\bf u} \otimes \phi^{\bf u}$  by means of the channel $T_{n}$.
We now seek to implement this in a physical situation.
The $N^{\bf u}$-part will follow in section
\ref{sec.energy}, the $\phi^{\bf u}$-part will be treated in this section.

We couple the $n$ qubits to a Bosonic field;
this is the physical implementation of LAN.
Subsequently, we perform a
measurement in the field  which will provide the information about the state
of the qubits; this is the utilization of LAN in order to
solve the asymptotic state estimation problem.

In this section we will limit ourselves to analyzing the joint
evolution of the qubits and field. The measurement on the field
is described in section
\ref{finmes}.

\subsection{Quantum stochastic differential equations}

In the weak coupling limit \cite{Gardiner} the joint evolution of the qubits and field can
be described mathematically by quantum stochastic differential equations (QSDE)
\cite{Hud.Par.}. The basic notions here are the Fock space, the creation and annihilation operators and the quantum stochastic differential equation of the unitary evolution. The Hilbert space of the field is the Fock space $\mathcal{F}(L^{2}(\mathbb{R}))$ as defined in \eqref{eq.Fock}. An important linearly complete set in
$\mathcal{F}(L^{2}(\mathbb{R}))$ is that of the exponential vectors
\begin{equation}\label{eq.exponential}
e(f) := \bigoplus_{n=0}^{\infty} \frac{1}{\sqrt{n!}} f^{\otimes n} :=\bigoplus_{n=0}^{\infty} \frac{1}{\sqrt{n!}} |f \rangle_{n}  , \qquad f\in L^{2}(\mathbb{R}),
\end{equation}
with inner product $\langle e(f), e(g)\rangle = \exp(\langle f, g\rangle)$.
The normalized exponential states
$\ket{f} := e^{-\langle f,f\rangle /2} e(f)$ are called
coherent states.
The vacuum vector is $|\Omega\rangle:=e(0)$ and we will denote the
corresponding density matrix $\ket{\Omega} \bra{\Omega}$ by $\Phi$.
The quantum noises are described by the creation and annihilation martingale 
operators
$A_{t}^{*}: = a^{*}(\chi_{[0,t]})$ and $A_{t}: = a(\chi_{[0,t]})$
respectively, where $\chi_{[0,t]}$ is the indicator function for $[0,t]$ and
$$
a(f) :e(g) \mapsto \langle f,g\rangle e(g).
$$
The increments $dA_{t} := a(\chi_{[0,t+dt]})-a(\chi_{[0,t]})$ and $dA^{*}_{t}$ 
play the role of non-commuting integrators in quantum stochastic differential equations, in the same way as the one can integrate against the Brownian motion in classical stochastic calculus.

We now consider the joint unitary evolution for qubits and field defined by the quantum stochastic differential equation \cite{Hud.Par.,Bouten&Guta&Maassen}:
$$
dU_{n}(t) = (a_{n}dA^{*}_{t} -a^{*}_{n} dA_{t} -\frac{1}{2} a^{*}_{n}a_{n} dt )U_{n}(t),
$$
where $U_{n}(t)$ is a unitary operator on $(\mathbb{C}^{2})^{\otimes n}\otimes \mathcal{F}(L^{2}(\mathbb{R}))$, and
$$
a_{n}: = \frac{1}{\sqrt{2j_{n}}}\sum_{k=1}^{n} \sigma_{+}^{(k)} ,\qquad
 \sigma_{+}^{(k)} := \mathbf{1}\otimes \dots \otimes (\sigma_{x} +i \sigma_{y})/2 \otimes \dots \otimes \mathbf{1}, \quad j_{n} := (\mu-1/2)n.
$$
As we will see later, the ``coupling factor'' $1/\sqrt{j_n}$ of the order
$n^{-1/2}$, is necessary in order to obtain convergence to the unitary evolution of the quantum harmonic oscillator and the field.

We remind the reader that the $n$-qubit space can be decomposed into irreducible representations as in \eqref{eq.decomposition}, and the interaction between the qubits and field respects this decomposition
$$
U_{n}(t) = \bigoplus_{j=0,1/2}^{n/2} U_{j,n}(t) \otimes \mathbf{1},
$$
where $\mathbf{1} $ is the identity operator on the multiplicity space $\mathcal{H}_{n}^{j}$, and
$$
U_{j,n}(t) : \mathcal{H}_{j}\otimes \mathcal{F}(L^{2}(\mathbb{R})) \to
\mathcal{H}_{j}\otimes \mathcal{F}(L^{2}(\mathbb{R})) ,
$$
is the restricted cocycle
\begin{equation}\label{eq.cocycle.j}
d U_{j,n}(t) = (a_{j}dA^{*}_{t} -a^{*}_{j} dA_{t} -\frac{1}{2} a^{*}_{j}a_{j} dt )U_{j,n}(t),
\end{equation}
with
$a_{j} $ acting on the basis $|j,m\rangle$ of $\mathcal{H}_{j}$ as
\begin{eqnarray*}
&&
a_{j} |j,m\rangle = \sqrt{j-m} \sqrt{ (j+m+1)/2j_{n}} \, |j,m+1\rangle ,\\
&&
a^{*}_{j} |j,m\rangle = \sqrt{j-m+1} \sqrt{ j+m/2j_{n}}  \,  |j,m-1 \rangle .
\end{eqnarray*}
{\bf Remark.} We point out that the {\it lowering} operator for $L_{z}$ acts as {\it creator} for our cut-off
oscillator since the highest vector $|j,j\rangle$ corresponds by $V_{j}$ to the vacuum of the oscillator. This choice does not have any physical meaning but is only related with our convention $\mu>1/2$. Had we chosen $\mu<1/2$, then the raising operator on the qubits would correspond to creation operator on the oscillator.

By \eqref{blocks} the initial state $\rho^{\otimes n}$ decomposes in the same way as the unitary cocycle, and thus the whole evolution decouples into separate ``blocks'' for each value of $j$. We do not have explicit solutions to these equations but based on the conclusions drawn from LAN we expect that 
as $n\to\infty$, the solutions will be well approximated by similar ones for a coupling between an oscillator and the field, at least  for the states in which we are interested. As a warm up exercise we will start with this simpler limit case where the states can be calculated explicitly.

\subsection{Solving the QSDE for the oscillator}\label{sec.oscillator.field}

Let $a^{*}$ and $a$ be the creation and annihilation operators of a quantum oscillator acting on $\mathcal{F}(\mathbb{C})$.  We couple the oscillator with the Bosonic field
and the joint unitary evolution is described by the family of unitary operators $U(t)$ satisfying the quantum stochastic differential equation
$$
dU(t) =
(adA^{*}_{t} -a^{*} dA_{t} -\frac{1}{2} a^{*}a dt )U(t).
$$
We choose the initial (un-normalized) state $\psi (0):= e({\bf z} )\otimes
|\Omega\rangle$, where ${\bf z}$ is any complex number, and we shall find the explicit form of the vector state of the system and field at time $t$:
$\psi(t):=U(t) \psi(0)$.

We make the following ansatz: $\psi(t) = e(\alpha_{t}) \otimes e(f_{t})$,
where $f_t(s) := f(s) \chi_{[0,t]}(s)$ for some $f \in L^2(\RR)$.
For each $\beta \in \CC$, $g \in L^2(\RR)$, define
$I(t) := \inp{e(\beta) \otimes e(g)}{\psi(t)}$.
We then have $I(t) = \exp(\bar{\beta} \alpha(t) + \inp{g}{f_t})$,
so that it satisfies
\begin{equation}\label{sarcofaag}
d I(t) = \left( \bar{\beta} \dt \alpha(t) +
\bar{g}(t) f(t) \right)I(t) dt\,.
\end{equation}

We now calculate $\frac{d}{dt} I(t)$ with the help of the QSDE.
Since $A_{t} e(f) = \inp{\chi_{[0,t]}}{f} e(f)$,
we have, for continuous $g$,
$dA_t e(g) = g(t)e(g) dt$.
However, since $A_{s} e(f_t)$ is constant for $s \geq t$,
we have $dA_t e(f_t) = 0$.
Thus
\begin{equation}\label{knaagdier}
d I(t) =
\inp{e(\beta) \otimes e(g)}{(adA^{*}_{t} -a^{*} dA_{t} -
\half a^{*}a dt ) \psi(t)} = (\bar{g}(t) \alpha(t)  -
\half \bar{\beta} \alpha(t)) I(t) dt\,.
\end{equation}

Equating \eqref{sarcofaag} with \eqref{knaagdier}
for all $t$, $\beta$ and continuous $g$,
we find $f(s) = \alpha(s)$,
$\frac{d}{dt} \alpha(t) = - \frac{1}{2} \alpha(t)$.
Thus $\alpha(t) = \alpha(0) e^{- \frac{1}{2} t}$,
$f_t(s) = \alpha(0)\chi_{[0,t]}(s) e^{- \frac{1}{2} s}$ with $\alpha(0)={\bf z}$.
 
In conclusion $\psi(t) = e({\bf z} e^{-\frac{1}{2}t}) \otimes
e({\bf z}  e^{-\frac{1}{2} s} \chi_{[0,t]}(s))$. For later use we denote the {\it normalized} solution by $\psi_{\bf z}(t):= U(t) |{\bf z}\rangle \otimes |\Omega\rangle = e^{-|{\bf z}|^{2}/2}U(t) e({\bf z})\otimes | \Omega \rangle$.

\subsection{QSDE for large spin}\label{sec.unitary.evol}

We consider now the unitary evolution for qubits and field:
$$
dU_{n}(t) = (a_{n}dA^{*}_{t} -a^{*}_{n} dA_{t} -\frac{1}{2} a^{*}_{n}a_{n} dt )U_{n}(t).
$$
It is no longer possible to obtain an explicit expression for the joint
vector state $\psi_{n}(t)$ at time $t$.
However we will show that for the states in which we are interested, a satisfactory explicit {\it approximate} solution exists.

The trick works for an arbitrary family of unitary solutions of a quantum stochastic differential equation $dU(t)= G_{dt}U(t)$, and the general idea is the following: if $\psi(t)$ is the true state $\psi(t) = U(t)\psi$ and $\xi(t)$ is a vector describing an approximate evolution ($\psi(0)=\xi(0)$) then with $U^{t}_{t+dt} :=U(t+dt)U(t)^{-1}$  we get
\begin{eqnarray*}
\psi(t+dt) - \xi(t+dt) &=& \psi(t+dt) - U^{t}_{t+dt} \xi(t) + U^{t}_{t+dt}\xi(t) -\xi(t) +\xi(t) -\xi(t+dt)\\
&=&
U^{t}_{t+dt} \left[ \psi(t) -\xi(t) \right] + [U(t+dt) - U(t)]U(t)^{-1} \xi(t) \\
&+& [\xi(t)  -\xi(t+dt)]\\
&=&
U^{t}_{t+dt} \left[ \psi(t) -\xi(t) \right] + G_{dt} \xi(t) - d\xi(t).
\end{eqnarray*}
By taking norms we get
\begin{equation}\label{eq.approx.qsde.solution}
d\| \psi(t) - \xi(t) \| \leq \| G_{dt} \xi(t) - d\xi(t)\|.
\end{equation}
The idea is now to devise a family $\xi(t)$ such that the right side is as small as possible.

We apply this technique block-wise, that is to each unitary $U_{j,n}(t)$ acting on
$\mathcal{H}_{j}\otimes \mathcal{F}(L^{2}(\mathbb{R}))$ (see equation \eqref{eq.cocycle.j})
for a ``typical'' $j\in \mathcal{J}_{n}$ (see equation  \eqref{eq.typical.j}).
By means of the isometry $V_{j}$ we can embed the space $\mathcal{H}_{j}$ into the first $2j+1$ levels of the oscillator and for simplicity we will keep the same notions as before for the operators acting on $\mathcal{F}(\mathbb{C})$. As initial states for the qubits we choose the block states $\rho^{\bf u}_{j,n}$.

\begin{theorem}\label{th.unitary.evolution}
Let $\rho^{\bf u}_{j,n} (t) = U_{j,n} (t)\, \left[ \rho^{\bf u}_{j,n} \otimes  \Phi\right] \, U_{j,n}^* (t)$ be the $j$-th block of the state of qubits and field at time $t$.
Let $\phi^{\bf u}(t):= U(t) \,\left[\phi^{\bf u} \otimes \Phi \right]\,
U(t)^* $ be the joint
state of the oscillator and field at time $t$.  For any $\eta <1/6$, for any
$\epsilon>0$,
\begin{equation}\label{eq.unitary.error}
 \sup_{j\in\mathcal{J}_{n}} \,
 \sup_{\|{\bf u} \| \leq n^{\eta}} \, \sup_{t} \| \rho^{\bf u}_{j,n} (t) -
\phi^{\bf u}(t) \|_{1} =O(n^{-1/4+\eta + \epsilon}, n^{-1/2+ 3\eta + \epsilon}).
\end{equation}
\end{theorem}

\noindent
{\it Proof.} From the proof of the local asymptotic normality Theorem \ref{th.qlan} we know that the initial states of the two unitary evolutions are asymptotically close to each other
\begin{equation}\label{eq.qlan.evolution}
\sup_{j\in\mathcal{J}_{n}} \,
 \sup_{\|{\bf u} \| \leq n^{\eta}} \| \rho^{\bf u}_{j,n}  - \phi^{\bf u}
\|_{1} = O(n^{-1/4+\eta + \epsilon}).
 \end{equation}

The proof consists of two estimation steps. In the first one, we will devise another initial state $\tilde{\rho}^{\bf u}_{j,n}$ which is an approximation of $\phi^{\bf u}$ and thus also of $\rho^{\bf u}_{j,n}$:
\begin{equation}\label{eq.error.initial.condition}
\sup_{j\in\mathcal{J}_{n}} \,
 \sup_{\|{\bf u} \| \leq n^{\eta}} \| \tilde{\rho}^{\bf u}_{j,n}  -
\phi^{\bf u}  \|_{1} = O( e^{-n^{\epsilon}}).
\end{equation}
In the second estimate we show that the evolved states $\tilde{\rho}^{\bf u}_{j,n}(t)$ and
$\phi^{\bf u}(t)$ are asymptotically close to each other
\begin{equation}\label{eq.error.evolved.states}
\sup_{j\in\mathcal{J}_{n}} \,
 \sup_{\|{\bf u} \| \leq n^{\eta}} \sup_{t} \| \tilde{\rho}^{\bf u}_{j,n}(t)  -
 \phi^{\bf u}(t)  \|_{1} =
 O(n^{-1/4+\eta +\epsilon}, n^{-1/2 + 3\eta + \epsilon}).
\end{equation}
This estimate is important because, the two trajectories are driven by different Hamiltonians, and in principle there is no reason why they should stay close to each other.

>From \eqref{eq.qlan.evolution}, \eqref{eq.error.initial.condition} and  \eqref{eq.error.evolved.states}, and using triangle inequality we get
\begin{equation*}
\sup_{j\in\mathcal{J}_{n}} \,
 \sup_{\|{\bf u} \| \leq n^{\eta}}
 \sup_{t}\| \rho^{\bf u}_{j,n} (t) - \phi^{\bf u}(t)  \|_{1} =
 O(n^{-1/4+\eta + \epsilon},
n^{-1/2 + 3\eta + \epsilon}).
\end{equation*}

The following diagram illustrates the above estimates. The upper line concerns the
time evolution of the block state $\rho^{\bf u}_{j,n}$ and the field. The lower
line describes the time evolution of the oscillator and the field. The estimates show that the diagram is ``asymptotically commutative'' for large $n$.
\begin{center}
$
\CD
\CS(\mathcal{H}_{j}) @> {\rm Id}_{j} \otimes \Phi>> \CS(\mathcal{H}_{j} \otimes \F) @>U_{j,n}(t)>>
\CS(\mathcal{H}_{j} \otimes \F) %@>>> \CS({\cal C})
 \\
@V {V_{j}\cdot V_{j}^{*} }VV  @VVV  @VVV % @ |
 \\
\CS(\F(\CC)) @>{\rm Id} \otimes \Phi >> \CS(\F(\CC) \otimes \F) @>U(t)>>
\CS(\F(\CC) \otimes \F) %@>>> \CS({\cal C})
\endCD
$
\end{center}
For the rest of the proof, we refer to Appendix \ref{kool}.

\qed

We have shown how the mathematical statement of LAN (the joint state of qubits converges to a Gaussian state of a quantum oscillator plus a classical Gaussian random variable) can in fact be physically implemented by coupling the spins to the environment and letting them ``leak'' into the field. In the next section, we will use this for the specific purpose of estimating $\bf u$ by performing a measurement in the field.

%\section{Heterodyne measurement}
%\label{finmes}

\section{The second stage measurement}
\label{finmes}

We now describe the second stage of our measurement procedure.
Recall that in the first stage a relatively small part $\tilde{n}=n^{1-\kappa}, 1>\kappa>0,$
of the qubits is measured and a rough estimator $\tilde{\rho}_{n}$ is obtained. The purpose of this estimator is to localize the state within a small neighborhood such
that the machinery of local asymptotic normality of Theorem \ref{th.qlan} can be applied.

In Theorem \ref{th.unitary.evolution} the local asymptotic normality was extended to the level of time evolution of the qubits interacting with a bosonic field. We have proven that at time $t$ the joint state of the qubits and field is
\begin{eqnarray*}
\rho_n^{\bf u}(t)
& := &\bigoplus_{j=0,1/2}^{n/2} p_{n,{\bf u}}(j) \frac1{2\pi s^2}\int_{ \mathbb{C} } d{\bf z}\,
e^{-|{\bf z} -
\sqrt{2\mu -1} \alpha_{\bf u}|^2/2s^2} \exp(-|{\bf z}|^2)
\times
\\& &
|e({\bf z} e^{-t/2 })_j\rangle\langle e({\bf z}
e^{-t/2 })_j| \otimes
|e({\bf z} e^{-u/2 }\chi_{[0,t]}(u))\rangle \langle e({\bf z} e^{-u/2 }\chi_{[0,t]}(u))|
\\& &
+ O(n^{\eta -1/4 +\epsilon }, n^{3\eta -1/2 + \epsilon}),
\end{eqnarray*}
for $\|{\bf u}\|\leq n^{\eta}$.
The index $j$ serves to remind the reader that the first exponential states
live in different copies $\mathcal{F}(\mathbb{C})_j$ of the oscillator space,
corresponding to $\mathcal{H}_j$ via the isometry $V_j$. We will continue to
identify $\mathcal{H}_j$ with its image in $\mathcal{F}(\mathbb{C})_j$.

We can now approximate the above state by its limit for large $t$, since
\begin{align}
\label{approx.time}
\exp(-|{\bf z}|^2) 
\langle e({\bf z} e^{-t/2 })_j  | \, j,j  \rangle 
\langle e({\bf z}
 e^{-u/2 }
\chi_{[0,t]}(u))\, | \,e({\bf z} e^{-u/2}) \rangle  =
\exp( - |{\bf z}|^2 e^{-t}).
\end{align}

As we are always working with $\|{\bf u}\|\leq n^{\eta}$, the only relevant
${\bf z}$
are bounded by $n^{\eta + \delta}$ for small $\delta$.
(The remainder of the Gaussian
integral has an exponentially decreasing norm, as discussed before).
Thus, for large enough time (i.e. for $t \geq \ln(n)$),
 we can write $\rho_n^{\bf u}(t) =
 \rho_n^{\bf u}(\infty)  + O(n^{\eta -1/4 + \epsilon} ,
 n^{3\eta -1/2 + \epsilon})$ with
\begin{eqnarray}
 \rho_n^{\bf u}(\infty) &:=&
 \bigoplus_{j=0,1/2}^{n/2} p_{n,{\bf u}}(j) |j, j\rangle\langle j,j|
\otimes \nonumber\\
& &
\left[ \frac1{2\pi s^2}\int_{ \mathbb{C} } d{\bf z} \, e^{-|{\bf z} -
\sqrt{2\mu -1} \alpha_{\bf u}|^2/2s^2}
|e({\bf z} e^{-u/2 })\rangle \langle e({\bf z} e^{-u/2})|
\exp(-|{\bf z}|^2) \right]. \label{limstate}
\end{eqnarray}

Thus, the field is approximately in the state 
$\phi^{\bf u}$ depending on $(u_{x}, u_{y})$, which is carried by the mode 
$(u\mapsto e^{-u/2} \chi_{[0,\infty)} (u))\in L^{2}(\mathbb{R})$ denoted for simplicity by $e^{-u/2}$. The atoms end up in a mixture of $|j,j\rangle$ states with coefficients 
$p_{n,{\bf u}}(j)$, which depend only on $u_{z}$, and are well approximated by the Gaussian random variable $N^{\bf u}$ as shown in Theorem \ref{th.qlan}.  
Moreover since there is no correlation between atoms and field, the statistical problem decouples into one concerning the estimation of the displacement in a family of Gaussian states $\phi^{\bf u}$, and one for estimating the center of $N^{\bf u}$.

For the former problem, the optimal estimation procedure is known to be the heterodyne measurement \cite{Holevo,Yuen&Lax}; for the latter, we perform a  ``which block'' measurement. These measurements are described in the next two subsections.

\subsection{The heterodyne measurement}
\label{subsec.hetero}

A heterodyne measurement is a ``joint measurement'' of the quadratures 
${\bf Q}:= (a+a^{*})/\sqrt{2}$ and ${\bf P}:= -i(a-a^{*})/\sqrt{2}$ of a quantum harmonic oscillator which in our case represents a mode of light. Since the two operators do not commute, the price to pay is the addition of some ``noise'' which will allow for an approximate measurement of both operators. The light beam passes through a beamsplitter having a vacuum mode as the second input, and then one performs a homodyne (quadrature) measurement on each of the two emerging beams. If ${\bf Q}_{v}$ and ${\bf P}_{v}$ are the vacuum quadratures then we measure the following output quadratures 
${\bf Q}_{1} := ({\bf Q} + {\bf Q}_{v})/\sqrt{2}$ and 
${\bf P}_{2} := ({\bf P} - {\bf P}_{v})/\sqrt{2}$, with $[{\bf Q}_{1}, {\bf P}_{2}]
=0$. Since the two input beams are independent,  the distribution of $\sqrt{2}{\bf Q}_{1}$ is the convolution between the distribution of ${\bf Q}$ and the distribution of ${\bf Q}_{v}$, and similarly for $\sqrt{2}{\bf P}_{2}$. 

In our case we are interested in the mode 
$e^{-u/2}$ which is in the state $\phi^{\bf u}$, up to a factor of order
$O(n^{\eta -1/4 + \epsilon} , n^{3\eta -1/2 + \epsilon})$. From \eqref{eq.displacedthermal} we obtain that the distribution of ${\bf Q}$ is $N(\sqrt{2(2\mu-1)} u_{x}, 1/(2(2\mu-1)))$, 
that of ${\bf P}$ is $N(\sqrt{2(2\mu-1)} u_{y}, 1/(2(2\mu-1)))$, and the joint distribution 
of the rescaled output 
$$
\left( ({\bf Q}+ {\bf Q}_{v})/ \sqrt{2(2\mu-1)} \, , \, ({\bf P}- {\bf P}_{v})/ \sqrt{2(2\mu-1)} \right) ,
$$
is 
\begin{equation}\label{gaus1}
N(u_{x} ,  \mu/(2(2\mu-1)^2)) \times N(u_{y} ,  \mu/(2(2\mu-1)^2)).
\end{equation}
We will denote by $(\tilde{u}_x, \tilde{u}_y)$ the result of the heterodyne measurement rescaled by the factor $\sqrt{2\mu-1}$ such that with good approximation 
$(\tilde{u}_x, \tilde{u}_y)$ has the above distribution and is an unbiased estimators of the parameters $(u_x, u_y)$.

Since we know in advance that the parameters $(u_{x}, u_{y})$ must be within the radius 
of validity of LAN we modify the estimators $(\tilde{u}_{x}, \tilde{u}_{y})$ 
to account for this information and obtain the final estimator $(\hat{u}_{x}, \hat{u}_{y})$:
\begin{align}
\label{mod1}
\hat{u}_i = \left\{\begin{array}{cc}
 \tilde u_i &\quad \textrm{if $|\tilde u_i|\leq 3 n^{\eta}  $} \\
  0 & \quad        \textrm{if $|\tilde u_i|>  3 n^{\eta}  $}
\end{array}
 \right.
\end{align}

Notice that if the true state $\rho$ is in the radius 
of validity of LAN around $\tilde \rho$, then $\|{\bf u}\|\leq n^{\eta}$, so that $|\hat{u}_i -
u_i |\leq |\tilde{u}_i - u_i|$. We shall use this when proving optimality of the
estimator. 

\subsection{Energy measurement}        
\label{sec.energy}

Having seen the $\phi^{\bf u}$-part,
we now move to the $N^{\bf u}$-part of the equivalence
between $\rho_{n}^{\bf u}$ and $N^{\bf u} \otimes \phi^{\bf u}$.
This too is a coupling to a bosonic field, albeit a different coupling.
We also describe the measurement in the field which will provide the
information on the qubit states.

The final state of the previous measurement, restricted to the atoms alone
(without the field), is obtained by a partial trace of
equation \eqref{limstate} (for large time) over the field
$$
\tau_{n}^{\bf u} = \sum_{j=0,1/2}^{n/2} p_{n,{\bf u}}(j) |j,j\rangle\langle j,j|
+ O(n^{\eta -1/4 + \epsilon} , n^{3\eta -1/2 + \epsilon}) \,.
$$
We will take this as the initial state of the second measurement, which will
\mbox{determine j.}

A direct coupling to the $J^2$ does not appear to be physically available,
but a coupling to the energy$J_z$ is realizable. This suffices, because the above state satisfies
$j = m$ (up to order $O(n^{\eta -1/4 + \epsilon}, n^{3\eta -1/2 + \epsilon})$).
We couple the atoms to a new field (in the vacuum state $\ket{\Omega}$)
by means of the interaction
$$
dU_t = \{ J_z (dA_{t}^{*} - dA_{t}) - \half J_{z}^2 dt \} U_t \,,
$$
with $J_z := \frac{1}{\sqrt{n}}\sum_{k=1}^{n}\sigma_z$.
Since this QSDE is `essentially commutative', i.e. driven by a single
classical noise
$B_t = (A^*_{t} - A_{t})/i$, the solution is easily seen to be
$$
U_t = \exp(J_z \otimes (A^*_{t} - A_t)) \, .
$$
Indeed, we have $d f(B_t) = f'(B_t)dB_t + \half f''(B_t)dt$ by the classical It\^o rule, so that
$$
d \exp (i J_z \otimes B_t) = \{ iJ_z dB_t - \half J_z^2 dt \}
\exp (iJ_z \otimes B_t)\, .
$$
For an initial state $\ket{j,m}\otimes \ket{\Omega}$, this evolution
gives rise to the final state
\begin{eqnarray*}
U_t \ket{j,m}\otimes\Omega &=& \ket{j,m}\otimes
\exp((m/\sqrt{n})(A^{*}_t - A_{t}))\Omega \\
&=& \ket{j,m}\otimes \ket{(m/\sqrt{n}) \chi_{[0,t]}},
\end{eqnarray*}
where $\ket{f} \in \mathcal{F}(L^2(\mathbb{R}))$ denotes the normalized
vector $\exp(-\langle f,f \rangle/2)e(f)$.
Applying this to the states $\ketbra{j,j}$ in $\tau_{n}^{\bf u}$
yields
$$
U_{t} \,\tau_{n}^{\bf u} \otimes \Phi\, U^*_{t} = \sum_{j=0,1/2}^{n/2} p_{n,{\bf u}}(j)
\ketbra{j,j} \otimes \ket{j/\sqrt{n} \chi_{[0,t]}} \bra{j/\sqrt{n} \chi_{[0,t]}}  
+ O(n^{\eta -1/4 + \epsilon}, n^{3\eta -1/2 + \epsilon})\,.
$$
The final state of the field results from a partial trace over the atoms;
it is given by
\begin{equation}\label{smeerpoets}
\sum_{j=0,1/2}^{n/2} p_{n,{\bf u}}(j) \,
\ketbra{ (j/\sqrt{n}) \chi_{[0,t]}}
+ O(n^{\eta -1/4 + \epsilon}, n^{3\eta -1/2 + \epsilon})\,.
\end{equation}

We now perform a homodyne measurement on the field, which amounts to a direct
measurement of $(A_t + A^*_t)/2t$.
In the state $\ket{(j/\sqrt{n} \chi_{[0,t]}}$, this yields
the value of $j$ with certainty for large time (i.e. $t\gg \sqrt{n}$).
Indeed, for this state, $\mathbb{E}((A_t + A^*_t)/2t) = j/\sqrt{n}$,
whereas $\mathbb{V}\mathrm{ar}(A_t + A^*_t)/2t) = 1/(4t)$.
Thus the probability distribution $p_{n,{\bf u}}$ is reproduced up to order
$O(n^{\eta -1/4 + \epsilon} , n^{3\eta -1/2 + \epsilon} )$ in $L^1$-distance.

The following is a remider from the proof of Theorem \ref{th.qlan}. 
If we start with $j$ distributed according to $p_{n}(j)$ and we smoothen $\frac{j}{\sqrt{n}} - \sqrt{n} (\mu-1/2)$ with a Gaussian kernel, then we obtain a random variable $g_n$ which is continuously distributed on $\mathbb{R}$ and converges in distribution to $N(u_z, \mu(1 - \mu))$, the error term being of order $O(n^{\eta - 1/2}) + O(n^{\epsilon - 1/2})$. For $j$ distributed according to the actual distribution, as measured by the
homodyne detection experiment, we can therefore state that $g_n$
is distributed according to
\begin{equation}
\label{gaus2}
N(u_z,\mu(1-\mu)) + O(n^{\eta -1/4 + \epsilon}, n^{3\eta -1/2 + \epsilon})
+ O(n^{\eta - 1/2}) + O(n^{\epsilon - 1/2}).
\end{equation}

As in the case of $(\hat{u}_{x}, \hat{u}_{y})$, we take into account the range of validity of LAN by defining the final estimator 
\begin{align}
\label{mod2}
\hat{u}_z = \left\{\begin{array}{cl}
 g_{n} &\quad \textrm{if $| g_{n}|\leq 3 n^{\eta}  $} \\
  0 & \quad        \textrm{if $| g_{n}|>  3 n^{\eta}  $\,.}
\end{array}
 \right.
\end{align}
Similarly, we note that if the true state $\rho$ is in the radius of validity of LAN around $\tilde \rho$, then $\|{\bf u}\|\leq n^{\eta}$, so that 
$|\hat{u}_z -u_z |\leq |\tilde{u}_z - u_z|$.

\section{Asymptotic optimality of the estimator}          \label{sec.endresult}

In order to estimate the qubit state,
we have proposed a strategy consisting of the following steps.
First, we use
$\tilde{n}:=n^{1-\kappa}$ copies of the state $\rho$ to get a rough estimate
$\tilde \rho_{n}$. Then we couple the remaining qubits with a field, and perform a
heterodyne measurement. Finally, we couple to a different field,
followed by homodyne measurement. From the measurement outcomes,
we construct an estimator $\hat{\rho}_{n} := \rho_{\hat{\bf u}_{n}/\sqrt{n}}$.

This strategy is asymptotically optimal in a global sense: for {\it any} true
state $\rho$ even if we knew beforehand that the true state $\rho$ is in a small
ball around a known state $\rho_0$, it would
be impossible to devise an estimator that could do better asymptotically, than our estimator $\hat{\rho}_{n}$ on a small ball around $\rho$. More precisely:
\begin{theorem}  \label{automobiel}
Let $\hat{\rho}_{n}$ be the estimator defined above.
For any qubit state $\rho_{0}$ different from the totally mixed state, for any sequence of
estimators $\hat{\varrho}_n$, the following local asymptotic minimax result
holds for any $0<\epsilon<1/12$:
\begin{equation}
\label{resultat}
\limsup_{n\to\infty}\sup_{\|\rho - \rho_{0}\|_1 \leq n^{-1/2+\epsilon}}
 n R(\rho,\hat{\rho}_{n}) \leq
 \limsup_{n\to\infty}\sup_{\|\rho - \rho_{0}\|_1 \leq n^{-1/2+\epsilon}}
n R(\rho,\hat{\varrho}_n).
\end{equation}
Let \mbox{$(\mu_{0}, 1-\mu_{0})$}  be the eigenvalues of $\rho_{0}$ with $\mu_{0}>1/2$. Then the local asymptotic minimax risk is
\begin{equation}
\label{risquefinal}
\limsup_{n\to\infty}\sup_{\|\rho - \rho_{0}\|_1 \leq n^{-1/2+\epsilon}}
n R(\rho,\hat{\rho}_{n}) = R_{\rm minimax}(\mu_{0}) =8\mu_{0}-4\mu_{0}^2  .
\end{equation}
\end{theorem}
\begin{proof}

We write the risk as the sum of two terms corresponding to the events
$E$ and $E^{c}$
that $\tilde{\rho}_{n}$ is inside or outside the ball of radius $n^{-1/2+\epsilon}$ around $\rho$. Recall that LAN is valid inside the ball. Thus
$$
R(\rho,\hat{\rho}_{n})  =
\mathbb{E} (\| \rho-\hat{\rho}_{n}\|_{1}^{2}\, \chi_{E^{c}} ) +
\mathbb{E} (\| \rho-\hat{\rho}_{n}\|_{1}^{2} \,\chi_{E} ),
$$
where the expectation comes from $\hat{\rho}_n$ being random. The distribution
of the result $\hat{rho_n}$ of our measurement procedure applied to the true unknown
state $\rho$ depends on $\rho$.
We bound the first part by $R_{1}$ and the second part by $R_{2}$ as shown below.

%The first contribution corresponds to the low probability event that our rough estimation
%$\tilde \rho_{n}$ is very wrong, so that $\rho$ is not in the radius of validity of quantum %local asymptotic normality.
$R_{1}$ equals $\mathbb{P}(E^{c})$ times the maximum error, which is $4$ since for any pair of density matrices $\rho$ and $\sigma$, we have $\| \rho-\sigma\|_{1}^{2}\leq 4$. Thus
$$
R_{1} = 4\mathbb{P} (\|\rho-\tilde{\rho}_{n} \|_{1} \geq n^{-1/2+\epsilon}).
$$
According to Lemma \ref{lemma.small.probability} this probability goes to zero exponentially fast, therefore the contribution brought by this term can be neglected.

We can now assume that $\tilde \rho_{n}$ is in the range of validity of local asymptotic normality and we can write $\rho^{\otimes n}= \rho^{\bf u}_{n}$ with ${\bf u}$ the
local parameter around $\tilde{\rho}_{n}$. We get the following inequalities for the second term in the risk.
\begin{eqnarray}
\mathbb{E} (\| \rho-\hat{\rho}_{n}\|_{1}^{2}\, \chi_{E} )
&\leq&
\mathbb{E} \left[ \|\hat{\rho}_{n} -\rho\|_{1}^{2} ~\Big|~ \|\tilde \rho_{n} - \rho\|_1
\leq n^{-1/2+\epsilon} ~\right] \nonumber \\
&\leq&
\sup_{ \|\rho - \rho_{0}\| < n^{-1/2+\epsilon}}
\mathbb{E} \left[ \|\hat{\rho}_{n} -\rho\|_{1}^{2} ~\Big|~ \tilde \rho_{n} =\rho_{0} \right]
\nonumber\\
&\leq&
\sup_{ \|\rho - \rho_{0}\| < n^{-1/2+\epsilon}}
\mathbb{E}_{\rho^{\bf u}_{n}(\infty)}
\left[ \|\hat{\rho}_{n} -\rho\|_{1}^{2} ~\Big|~ \tilde \rho_{n} =\rho_{0} \right]
 \nonumber \\
&+&
\sup_{ \|\rho - \rho_{0}\| < n^{-1/2+\epsilon}}
  \|\rho^{\bf u}_{n} (t) - \rho^{\bf u}_{n}(\infty)\|_{1} ~
  \sup_{ \hat{\bf u}_{n} } \|\hat{\rho}_{n} -\rho\|_{1}^{2}
  \nonumber    \\
&\leq&
\sup_{ \|\rho - \rho_{0}\| < n^{-1/2+\epsilon}}  \mathbb{E}_{\rho^{\bf u}_{n}(\infty)}
\left[  \|\hat{\rho}_{n} -\rho\|_{1}^{2} ~\Big|~ \tilde \rho_{n} =\rho_{0}   \right]
\nonumber  \\
&+&
 c n^{-1+2\eta}\sup_{ \|\rho - \rho_{0}\| < n^{-1/2+\epsilon}}
 \|\rho_{n}^{\bf u}(t) - \rho^{\bf u}_{n}(\infty) \|_1
 = R_{2}.
 \label{error}
\end{eqnarray}
The first  two inequalities are trivial. In the third inequality we change the expectation from the one with respect to the probability distribution of our data
$\mathbb{P}_{\rho^{\bf u}_{n}(t)}$ to the probability distribution
$\mathbb{P}_{\rho^{\bf u}_{n}(\infty)}$. In doing so, an additional term
$\|\mathbb{P}_{\rho^{\bf u}_{n}(t)} - \mathbb{P}_{\rho^{\bf u}_{n}(\infty)}\|_{1}$ appears which is bounded from above by  $\|\rho^{\bf u}_{n} (t) - \rho^{\bf u}_{n}(\infty)\|_{1}$.
In the last inequality we can bound $ \|\hat{\rho}_{n} -\rho\|_{1}^{2}$ by
$c n^{-1+2\eta}$ for some constant $c$. Indeed from definitions \eqref{mod1} and \eqref{mod2} we know that $\| \hat{\rho}_{n} -\rho_{0} \|_{1} \leq c^{\prime} n^{-1/2 +\eta} $ and additionally we are under the assumption 
$\| \rho -\rho_{0}\|_{1}\leq  n^{-1/2 +\epsilon}$ with $\epsilon<\eta$.

For the following, recall that all our LAN estimates are valid uniformly
around any state $\rho^{\bf 0}=\tilde{\rho}$ as long as $\mu-1/2\geq \epsilon_2 > 0$. 
As we are working with $\rho$ different
from the totally mixed state and $\|\rho -\tilde \rho\| \leq
n^{-1/2+\epsilon}$, we know that for  big enough $n$, $\tilde \mu - 1/2\geq
\epsilon_2$ for any possible $\tilde \rho$. We can then apply the uniform results of the previous sections.

The second term in $R_{2}$ is $O(n^{-5/4+ 3\eta +\delta}  ,n^{-3/2+5\eta+\delta })$ where $\delta>0$ can be chosen arbitrarily small. Indeed in the end of section  \ref{sec.timevolf} we have proven that after time $t\geq \ln n$,  the following holds:
$  \|\rho_{n}^{\bf u}(t) -\rho^{\bf u}_{n} (\infty) \|_1 =  O(n^{-1/4+\eta+ \delta}, n^{-1/2+ 3\eta+\delta} )$.
The contribution to $nR(\rho,\hat{\rho}_{n})$ brought by this term will not count in the 
limit, as long as $\eta$ and $\epsilon$ are chose such that $1/12 > \eta>\epsilon$.

We now deal with the first term in $R_{2}$. We write $\rho$ in local parametrization around $\rho_{0}=\tilde{\rho}$ as $\rho_{{\bf u}_n/\sqrt{n}}$. We have
\begin{eqnarray}
\label{eq.dist.param}
\|\hat{\rho}_{n}-\rho\|_{1}^{2}=
 \| \rho_{{\bf u}/\sqrt{n}} - \rho_{{\bf \hat{u}_{n}}/\sqrt{n}}\|_1^2 &=&
 4\frac{(u_z-\hat{u}_z)^2 + (2\mu-1)^2 ((u_x-\hat{u}_x)^2 +
(u_y-\hat{u}_y)^2)}{n} \notag\\
&+& O(\|{\bf u} - \hat{\bf u}_{n} \|^3 n^{-3/2}).
\end{eqnarray}
The remainder term
$O(\|{\bf u} - \hat{\bf u}_{n}\|^3 n^{-3/2})$ is negligible.
It is $O(n^{3\eta-3/2})$ which does not contribute to $nR(\rho,\hat{\rho}_{n})$ for $\eta<1/6$. This is because on the one hand we have asked for $
\|\tilde \rho_{n} - \rho\| < n^{-1/2+\epsilon} $, and on the other hand, we have
bounded our estimator $\hat{\bf u}_{n}$ by using \eqref{mod1} and \eqref{mod2}.

We now evaluate $\mathbb{E}_{\rho^{\bf u}_{n}(\infty)}
\left[ d({\bf u},\hat{\bf u}_{n})^{2} \right]$
with the notation
\begin{equation}\label{eq.loss.function}
d({\bf u}, {\bf v})^{2} := 4\left[ (u_z- v_z)^2 + (2\mu-1)^2 ((u_x- v_x)^2 +
(u_y- v_y)^2)\right] .
\end{equation}
Note that the risk of $\hat{\bf u}_{n}$ is smaller than that of $\tilde{\bf u}_{n}$
(see discussion below \eqref{mod1} and \eqref{mod2}). Under the law 
$\mathbb{P}_{\rho^{\bf u}_{n}(\infty)}$ the estimator  $\tilde{\bf u}_{n}$ has a Gaussian distribution  as shown in \eqref{gaus1} and \eqref{gaus2} with fixed and known variance and unknown expectation. In statistics this type of model is known as a Gaussian shift experiment \cite{vanderVaart}. Using
\eqref{gaus1} and \eqref{gaus2}, we get
$\mathbb{E}_{\rho^{\bf u}_{n}(\infty)}\left[ (u_z - \hat{u}_z)^2\right] \leq
\mu(1-\mu)$ and $\mathbb{E}_{\rho^{\bf u}_{n}(\infty)}\left[ (u_i -
\hat{u}_i)^2\right] \leq
\mu/(2(2\mu-1)^2) $ for $i=x,y$. Substituting these bounds in \eqref{eq.dist.param},
we obtain \eqref{risquefinal}.

\smallskip

We will now show that the sequence $\hat{\rho}_{n}$ is optimal in the local minimax sense: for any $\rho_{0}$ and any other sequence of estimators $\hat{\varrho}_{n}$ we have
$$
R_{0} =
\limsup_{n\to \infty} \sup_{\|\rho -\rho_{0}\|_{1} \leq n^{-1/2+\epsilon}} nR(\rho, \hat \varrho_{n}) \geq
8\mu_{0} -4\mu_{0}^{2}.
$$
We will first prove that the right hand side is the  minimax risk 
$R_{\rm minimax} (\mu_{0})$ for the family of states 
$N^{\bf u}\otimes \phi^{\bf u}$ which is the limit of the local families $\rho^{\bf u}_{n}$ 
of qubit states centered around $\rho_{0}$. We then extend the result to our sequence of quantum statistical models $\rho^{\bf u}_n$.

The minimax optimality for $N^{\bf u}\otimes \phi^{\bf u}$ can be checked separately for
the classical and the quantum part of the experiment. For the quantum part
$\phi^{\bf u}$, the optimal measurement is known to be the heterodyne measurement.
A proof of this fact can be found in Lemma 7.4 of \cite{Guta&Kahn}. 
For the classical part, which corresponds to the measurement of $L_{z}$, the
optimal estimator is simply the random variable $X\sim N^{\bf u}$ itself \cite{vanderVaart}. 

%Note that the variance of the 
%states $N^{\bf u}\otimes \phi^{\bf u}$ depends on the parameter $\mu_{0}$ which until now was mostly considered to be fixed. 
%In the following argument however we will use two different such families and for this purpose we will make the dependence on $\mu$ explicit by writing $N^{\bf u}_{\mu}\otimes \phi^{\bf u}_{\mu}$.

We now end the proof by using the other direction of LAN.
Suppose that there exists a better sequence of estimators $\hat{\varrho}_n$ such that
$$
R_{0}  < R_{\rm minimax} (\mu_{0}) = 8\mu_{0} - 4\mu_{0}^{2}.
$$ 
We will show that this leads to an estimator $\hat{u}$ of ${\bf u}$ for the family 
$N^{\bf u}\otimes \phi^{\bf u}$ whose maximum risk  is smaller than the  minimax risk $R_{\rm minimax}(\mu_{0})$, which is impossible. 
%The variance parameter 
%$\mu$ of the family will be chosen such that $\mu> \mu_{0}$ as we will see later.

By means of a beamsplitter one can divide the state 
$\phi^{\bf u}$ into two independent Gaussian modes, using a thermal state $\phi:=\phi^{0}$ as the second input. If $r$ and $t$ are the reflectivity and respective transmitivity of the beamsplitter ($r^{2}+t^{2}=1$), then the transmitted beam has state 
$\phi^{\bf u}_{tr} = \phi^{t{\bf u}}$ and the reflected one 
$\phi^{\bf u}_{ref} = \phi^{r{\bf u}}$.
%Then the state of the transmitted beam $\phi^{\bf u}_{tr}$ 
%is very close to the initial state $\phi^{\bf u}$ itself, and the reflected one is close to the thermal state and carries very little information about $\bf{u}$. 
By performing a heterodyne measurement on the latter, and observing the classical part $N^{\bf u}$, we can localize ${\bf u}$ within a big ball around the result $\tilde{\bf u}$ with high probability, in the spirit of Lemma \ref{lemma.small.probability}. 
More precisely, for any small $\tilde\epsilon>0$ we can find $a>0$ big enough such that the risk contribution from unlikely $\tilde{\bf u}$'s is small
$$
\mathbb{E}(\|{\bf u} - \tilde{\bf u}\|^{2} \chi_{ \|{\bf u} - \tilde{\bf u}\|>a})  < \tilde\epsilon.
$$
Summarizing the localization step, we may assume that the parameter ${\bf u}$ satisfies $\|{\bf u}\| < a$ 
with an $\tilde{\epsilon}$ loss of risk, where $a= a(r,\tilde{\epsilon})$.

Now let $n$ be large enough such that $n^{\epsilon}> a$, then the parameter ${\bf u}$ falls within the domain of convergence of the inverse map $S_{n}$ of 
Theorem \ref{th.qlan} and by \eqref{eq.channel.conv.inverse} 
(with $\epsilon$ replacing $\eta$ and $\delta$ replacing $\epsilon$) we have 
$$
\| \rho^{t{\bf u}}_{n} - S(N^{t{\bf u}} \otimes \phi^{t{\bf u}} )\|_{1} \leq C n^{-1/4+\epsilon+ \delta},
$$
for some constant $C$.

Next we perform the measurement leading to the estimator $\hat{\varrho}_n$ and equivalently to an estimator $\hat{\bf u}_{n}$ of 
${\bf u}$. Without loss of risk we can implement the condition 
$\|{\bf u}\|<a$ into the estimator  $\hat{\bf u}_{n}$ in a similar fashion as in 
\eqref{mod1} and \eqref{mod2}. 
The risk of this estimation procedure for $\phi^{\bf u}$ is then bounded 
from above by the sum of three terms: the risk $n R_{\rho}(\hat{\varrho}_{n})/t^{2} $ coming from the qubit estimation, the error contribution from the map $S_{n}$ which is  
$a^{2} n^{-1/4 + \epsilon +\delta}$, and the localization risk contribution 
$\tilde{\epsilon}$. This risk bound uses the same technique as the third inequality of \eqref{error}. The second contribution can be made arbitrarily small by choosing $n$ large enough, for $\epsilon<1/4$. From our assumption we have $R_{0} < R_{minimax}(\mu_{0})$ and we can choose $t$ close to one such that $R_{0} /t^{2}< R_{minimax}(\mu_{0})$ and further choose $\tilde{\epsilon}$ such that  
$R_{0} /t^{2}+\tilde{\epsilon}< R_{minimax}(\mu_{0})$.

In conclusion, we get that the risk for 
estimating ${\bf u}$ is asymptotically smaller that the risk of the heterodyne measurement combined with observing the classical part which is known to be 
minimax \cite{Guta&Kahn}. Hence no such sequence $\hat{\varrho}_{n}$ exists, and $\hat{\rho}_{n}$ is optimal.

\end{proof}

\noindent{\bf Remark.}
In Theorem \ref{resultat}, we have used the risk function
$R(\rho , \hat{\rho}) = \mathbb{E}(d^2(\rho , \hat{\rho})) $,
with $d$ the $L_1$-distance
$d(\rho , \hat{\rho}) = \|\rho - \hat{\rho}\|_{1}$.
However,
the obtained results can easily
%essentially\footnote{The heterodyne measurement may
%have to be adapted.} be applied
be adapted to \emph{any}
distance measure $d^2 (\rho_{\hat{\bf u}} , \rho_{\bf u} )$ which is locally
quadratic in $\hat{\bf u} - \bf{u}$, i.e.
$$
d^2(\rho_{\hat{\bf u}} , \rho_{\bf u} ) =
\sum_{\alpha, \beta = x,y,z} \gamma_{\alpha \beta}
(u_{\alpha} - \hat{u}_{\alpha}) (u_{\beta} - \hat{u}_{\beta})
+ O(\|u - \hat{u}\|^{3})\,.
$$

For instance, one may choose
$d^2(\hat{\rho} , \rho) = 1 - F^2(\hat{\rho} , \rho)$
with the fidelity $F(\hat{\rho} , \rho) :=
\mathrm{Tr}(\sqrt{\sqrt{\hat{\rho}} \rho \sqrt{\hat{\rho}}  })$.
For non-pure states, this is easily seen to be locally quadratic with
$$
\gamma =
\left(
\begin{array}{c c c}
(2 \mu_0 - 1)^2&0&0\\
0&(2 \mu_0 - 1)^2&0\\
0&0&\frac1{1 -(2 \mu_0 - 1)^2}\\
\end{array}
\right)     \,.
$$
For the corresponding risk function
$R_{F}(\rho,\hat{\rho}_{n}) :=
\mathbb{E}(1 - F^2(\rho,\hat{\rho}_{n}))$,
this yields
\begin{equation}
\label{risqueF}
\limsup_{n\to\infty}\sup_{\|\rho - \rho_{0}\|_1 \leq n^{-1/2+\epsilon}}
n R_{F}(\rho,\hat{\rho}_{n}) = \mu_0 + 1/4 \,,
\end{equation}
with the same asymptotically optimal $\hat{\rho}$.
The asymptotic rate $R_{F} \sim \frac{4\mu_0 + 1}{4n}$ was found earlier in
\cite{Bagan&Gill}, using different methods.

\section{Conclusions}

In this article, we have shown two properties of
quantum local asymptotic normality (LAN) for qubits.
First of all, we have seen that its radius
of validity is arbitrarily close to $n^{-1/4}$ rather than $n^{-1/2}$.
And secondly, we have seen how LAN
can be implemented physically, in a quantum optical setup.

We use these properties to construct an
asymptotically optimal estimator
$\hat{\rho}_n$ of the qubit state
$\rho$, provided that we are given $n$ identical copies
of $\rho$.
Compared with other optimal estimation methods
\cite{Bagan&Gill,Hayashi&Matsumoto}, our measurement technique
makes a significant step in the direction  of an experimental implementation.

%
%Physical implementation of this estimator appears to
%be within the reach of current technology.
The construction and optimality of $\hat{\rho}_n$
are shown in three steps.
\begin{itemize}
\item[I]
In the preliminary stage, we perform
measurements of $\sigma_x$, $\sigma_y$ and $\sigma_z$
on a fraction
$\tilde{n} = n^{1-\kappa}$ of the $n$ atoms.
As shown in section \ref{sec.estimation},
this yields a rough estimate $\tilde{\rho}_{n}$
which lies within a distance $n^{-1/2 + \epsilon}$
of the true state $\rho$ with high probability.
%($\eta$ must be slightly larger than $\epsilon/2$.)
\item[II]
In section \ref{sec.lan}, it is shown that
local asymptotic normality holds within a
ball of radius $n^{-1/2 + \eta}$ around $\rho$ ($\eta>\epsilon$).
This means that locally, for $n \rightarrow \infty$,
all statistical problems concerning the $n$
identically prepared qubits are equivalent
to statistical problems concerning a
Gaussian distribution $N^{\bf u}$
and its quantum analogue,
a displaced thermal state $\phi^{\bf u}$ of the harmonic oscillator.
\end{itemize}

Together, I and II imply that the principle of
LAN %quantum local asymptotic normality
has been extended to a global setting.
It can now be used for
a wide range of asymptotic statistical problems,
including the global problem of state estimation.
Note that this hinges on the rather subtle extension
of the range of validity of LAN to neighborhoods
of radius larger than $n^{-1/2}$.

\begin{itemize}
\item[III] LAN provides an abstract equivalence between
the n-qubit states $\rho^{\otimes n}_{{\bf u}/\sqrt{n}}$
on the one hand, and on the other hand
the Gaussian states $N^{\bf u} \otimes \phi^{\bf u}$.
In sections \ref{sec.timevolf} and \ref{finmes} it is shown that this abstract equivalence
can be implemented physically by two consecutive couplings
to the electromagnetic field.
For the particular problem of state estimation,
homodyne and heterodyne detection on the
electromagnetic field then yield the data from 
which the optimal estimator $\hat{\rho}_{n}$ is computed.
\end{itemize}

Finally, in section \ref{sec.endresult}, it is shown that
the estimator $\hat{\rho}_n$, constructed above,
is optimal in a local minimax sense.
Local here means that optimality holds
in a ball of radius slightly bigger than $n^{-1/2}$ around
\emph{any} state $\rho_0$ except the tracial state. That is,
even if we had known
beforehand that the true state
lies within this ball around $\rho_0$,
we would not have been able
to construct a better estimator than $\hat{\rho}_n$,
which is of course
%Of course, $\hat{\rho}_n$ is
independent of $\rho_0$.

For this asymptotically optimal estimator, we have shown that
the risk $R$ converges to zero at rate
$R(\rho,\hat{\rho}_{n}) \sim \frac{8\mu_{0}-4\mu_{0}^2}{n}$,
with $\mu_0 > 1/2$ an eigenvalue of $\rho$.
More precisely, we have
$$
\limsup_{n\to\infty}\sup_{\|\rho - \rho_{0}\|_1 \leq n^{-1/2+\epsilon}}
n R(\rho,\hat{\rho}_{n}) = 8\mu_{0}-4\mu_{0}^2 .
$$
The risk is defined as
$R(\rho,\hat{\rho}) = \mathbb{E}(d^2(\rho,\hat{\rho}))$,
where we have chosen
$d(\hat{\rho} , \rho)$ to be the
$L_1$-distance $\|\hat{\rho} - \rho\|_{1} := \mathrm{Tr}(|\hat{\rho} - \rho|)$.
This seems to be a rather natural choice because of its
direct physical significance
%$\|\hat{\rho} , \rho\|_{1}$ is
%proportional to
as the worst case difference between the probabilities
induced by $\hat{\rho}$ and $\rho$ on a single event.

Even still, we emphasize that the same procedure can be applied to a
wide range of other risk functions.
Due to the local nature
of the estimator $\hat{\rho}_n$ for large $n$,
its rate of convergence in a risk $R$
is only sensitive
to the lowest order Taylor expansion of $R$
in local parameters $\hat{\bf u} - \bf{u}$.
The procedure can therefore easily be adapted to
other risk functions, provided that the distance measure
$d^2 (\rho_{\hat{\bf u}} , \rho_{\bf u} )$
is locally
quadratic in $\hat{\bf u} - \bf{u}$.

\noindent
{\bf Remark.} The totally mixed state ($\mu=1/2$) is a singular point in the parameter space, and Theorem \ref{th.qlan} does not apply in this case. The effect of the 
singularity is that the family of states \eqref{eq.displacedthermal} collapses to a 
single degenerate state of infinite temperature. However this phenomenon is only due 
to our particular parametrisation, which was chosen for its convenience in 
describing the local neighborhoods around arbitrary states, with the exception of the totally mixed state. Had we chosen a different parametrisation, e.g. in terms of the
Bloch vector, we would have found that local asymptotic normality holds for the totally mixed state as well, but the limit experiment is different: it consists of a three dimensional 
{\it classical} Gaussian shift, each independent component corresponding to the local change in the Bloch vector along the three possible directions. Mathematically, the optimal measurement strategy in this case is just to observe the classical variables. However this strategy cannot be implemented by coupling with the field since this coupling becomes singular (see equation \eqref{eq.cocycle.j}). 

These issues become more important for higher dimensional systems where the eigenvalues may exhibit more complicated multiplicities, and will be dealt with in that context.

\smallskip

\begin{center}
{\bf Acknowledgments}
\end{center}
We thank Richard Gill for the many discussions which helped shape up the paper. 
We thank the School of Mathematics of the University of Nottingham,
as well as the Department of Mathematics of the University of Nijmegen,
for their warm hospitality during the writing of this paper. 
M. G. acknowledges the financial support received
from the Netherlands Organisation for Scientific Research (NWO).

\appendix

\section{Appendix: Proof of Theorem \ref{th.qlan}}\label{sec.proof.qlan}

Here we give the technical details of the proof of local asymptotic normality
with ``slowly growing'' local neighborhoods $\| {\bf u}\| \leq
n^{\eta}$, with $\eta<1/4$.
We start with the map $T_n$.

\subsection{Proof of Theorem \ref{th.qlan}; the map $T_n$}

 Let us define, for $0 < \epsilon < (1/4 -\eta) $ the interval
\begin{equation}\label{eq.typical.j}
 \mathcal{J}_{n} = \left\{ j \,:\, (\mu-1/2)n -n^{1/2+\epsilon}\leq j\leq (\mu-1/2)n +
n^{1/2+\epsilon}\right\}.
\end{equation}

Notice that $j\in \mathcal{J}_n$ satisfies $2j\geq \epsilon_{2} n$ for all
$\mu-1/2\geq \epsilon_2$ and $n$ big enough, independently of $\mu$.

Then $ \mathcal{J}_n$ contains the relevant values of $j$, uniformly for
$\mu-1/2\geq \epsilon_2$:
\begin{equation}
\label{tail}
\lim_{n\to\infty} p_{n,{\bf u}}(\mathcal{J}_{n})=1 - O(n^{-1/2+\epsilon}).
\end{equation}
This is a consequence of Hoeffding's inequality applied to the binomial distribution, and
recalling that $p_{n,{\bf u}}(j) = B(n/2 + j) (1 + O(n^{-1/2+\epsilon}))$ for
$j\in \mathcal{J}_n$.

We upper-bound $ \| T_n(\rho_n^{\bf u}) -N^{\bf u}\otimes \phi^{\bf u}\|$ by the sum
\begin{eqnarray}
&&
 3
\sum_{j\not\in \mathcal{J}_{n}} p_{n,j}^{\bf u}
+
\left \| N^{\bf u}  - \sum_{j\in\mathcal{J}_{n} } p_{n, {\bf u}}(j) \tau_{n,j}     \right\|_{1}
+
\sup_{j\in \mathcal{J}_n} \| V_{j} \rho_{j,n}^{\bf u} V_{j}^{*} - \phi^{\bf u}\|_{1}.
\end{eqnarray}

The first two terms are ``classical'' and converge to zero uniformly over $\|
{\bf u}\| \leq n^{\eta}$: for the first term, this is (\ref{tail}), while the
second term converges uniformly on $\mu-1/2\geq \epsilon_2$ at rate $n^{\eta-1/2}$ \cite{Kahn}. The third term can be analyzed as in Proposition 5.1 of \cite{Guta&Kahn}:
\begin{eqnarray}\label{eq.two.terms}
\left\|V_{j} \rho^{\bf u}_{n,j} V_{j}^{*} - \phi^{\bf u}\right\|_{1} & \leq &
\left\| \rho^{\bf u}_{n,j} - V_{j}^{*} \phi^{\bf u}V_{j}\right\|_{1} +
\left\| \phi^{\bf u} - P_{j} \phi^{\bf u} P_{j} \right\|_{1},
\end{eqnarray}
where $P_{j} := V_{j} V_{j}^{*}$ is the projection onto the image of
$V_{j}$. We will show that both terms on the right side go to zero uniformly at
rate $n^{-1/4+\eta+\epsilon}$ over
$j\in \mathcal{J}_{n}$ and
$\| {\bf u}\| \leq n^{\eta}$. The trick is to note that displaced thermal equilibrium states are Gaussian mixtures of coherent states
\begin{equation}\label{eq.displaced.thermalstate.coherent}
\phi^{{\bf u}} = \frac{1}{\sqrt{2\pi s^2}}
\int e^{- |{\bf z}-\sqrt{2\mu -1}\alpha_{\bf u}|^{2}/2s^{2}}
\left( | {\bf z} \rangle \langle {\bf z} | \right) d^{2} {\bf z},
\end{equation}
where $s^{2}:= (1-\mu)/(4\mu-2)$.

The second term on the left side of \eqref{eq.two.terms} is bounded from above by
$$
 \frac{1}{\sqrt{2\pi s^2}}
\int e^{- |{\bf z}-\sqrt{2\mu -1}\alpha_{\bf u}|^{2}/2s^{2}}
\|  | {\bf z} \rangle \langle {\bf z} |  - P_{j}  | {\bf z} \rangle \langle {\bf z} | P_{j} \|_{1} \,
d^{2} {\bf z},
$$
which after some simple computations can be reduced (up to a constant) to
\begin{equation}\label{zwaardvis}
\int e^{- |{\bf z}|^{2}/2s^{2}}
\| P_{j}^{\perp} |{\bf z}+\sqrt{2\mu -1}\alpha_{\bf u}\rangle\| \,
d^{2} {\bf z}.
\end{equation}

We now split the integral. the first part is integrating over $|{\bf z}|\geq
n^{\eta+\delta} $ with $0<\delta <1/4 -\eta/2 $. The integral is dominated by the Gaussian and its value is $O(e^{-n^{2(\eta + \delta)}/(2s^2)}) $.
The other part is bounded by the supremum over
$|{\bf z}|\leq 2n^{\eta+ \delta}$ (as $\|{\bf u}\|\leq n^{\eta}$) of $\|
P_{j}^{\perp} |{\bf z}\rangle\| $. Now $\| P_{j}^{\perp} |{\bf z}\rangle\| \leq
|{\bf z}|^{j}/\sqrt{j!} =O(e^{-  n (1/2-\eta- 2\delta) })$ uniformly on $j\in \mathcal{J}_n$,
for any
$\mu-1/2\geq \epsilon_2$ since then $2 j\geq \epsilon_2 n$.

The same type of estimates apply to the first term
\begin{eqnarray}
&&
\left\| \rho^{\bf u}_{n,j} -  V_{j}^{*}\phi^{\bf u}V_{j} \right\|_{1} =
\left\|
\mathrm{Ad} \left[ U_{j} \left(\frac{\bf u}{\sqrt{n}}\right) \right]
\left( \rho^{\bf 0}_{n,j} \right)
-V_{j}^{*}  \phi^{\bf u} V_{j}\right\|_{1}\leq \nonumber \\
&&
\left\|\rho^{\bf 0}_{n,j} - V_{j}^{*} \phi^{\bf 0} V_{j} \right\|_{1} +
\left\|
 \mathrm{Ad} \left[ U_{j} \left(\frac{\bf u}{\sqrt{n}}\right) \right]
\left( V^{*}_{j} \phi^{\bf 0} V_{j}\right)
-V_{j}^{*} \phi^{\bf u} V_{j}
\right\|_{1} .\label{eq.triangle.ineq.}
\end{eqnarray}
The first term on the right side does not depend on ${\bf u}$. From the proof
of Lemma 5.4 of \cite{Guta&Kahn}, we know that
\begin{align*}
\left\|\rho^{\bf 0}_{n,j} - V_{j}^{*} \phi^{\bf 0} V_{j} \right\|_{1} \leq
\left(\frac{p^{2j+1}}{1-p^{2j+1}}+ p^{2j+1}\right)
\end{align*}
with $p= (1-\mu)/\mu$. Now the left side is of the order $p^{2j+1}$ which
converges exponentially fast to zero uniformly on $\mu-1/2\geq \epsilon_2$ and $j\in
\mathcal{J}_{n}$.

The second term of \eqref{eq.triangle.ineq.} can be bounded again by a Gaussian integral
\begin{eqnarray}\label{eq.gaussian.integral}
&&
\frac{1}{\sqrt{2\pi s^{2}}}
\int e^{- |{\bf z}|^{2}/2s^{2}} \| \Delta({\bf u}, {\bf z}, j) \|_{1} d^{2}{\bf z},
\end{eqnarray}
where the operator $\Delta({\bf u}, {\bf z}, j) $ is given by
\begin{equation*}
\Delta({\bf u}, {\bf z}, j) :=
\mathrm{Ad} \left[ U_{j} \left({\bf u}/\sqrt{n}\right) \right]
\left( V^{*}_{j} |{\bf z}\rangle \langle{\bf z} | V_{j}\right)
 -  V_{j}^{*} \,
| {\bf z} + \sqrt{2\mu - 1}\alpha_{\bf u} \rangle \langle  {\bf z} + \sqrt{2\mu -1}\alpha_{\bf u} |  \,V_{j}.
\end{equation*}

Again, we split the integral along $\|{\bf z}\|\geq n^{\eta + \delta}$. The outer
part converges to zero faster than any power of $n$, as we have already seen.
The inner integral, on the other hand, can be bounded uniformly over $\|{\bf
u}\|\leq n^{\eta}$, $\mu-1/2\geq \epsilon_2$  and $j\in \mathcal{J}_n$ by the supremum of $\| \Delta({\bf
u}, {\bf z}, j)\|_1$  over
$|{\bf z}|\leq 2 n^{\eta + \delta}$, $\mu-1/2\geq \epsilon_2$,  $j\in \mathcal{J}_n$ and $\|{\bf u}\|\leq n^{\eta}$.

Let $\tilde{\bf z}\in\mathbb{R}^{2}$ be such that $\alpha_{\tilde{\bf z}}= {\bf z}/\sqrt{2\mu-1}$, and  denote $\psi(n,j, {\bf v}) =  V_{j}U_{j}( {\bf v}/\sqrt{n}) |j,j\rangle$. Then, up to a $\sqrt{2}$ factor,  $ \| \Delta({\bf u}, {\bf z}, j) \|_{1}$ is bounded from above by the
\begin{eqnarray}
&&
 \left\|
\psi(n,j, \tilde{\bf z}) -
|{\bf z}\rangle
\right\|+ \nonumber\\
&&
\left\|
\psi(n,j, {\bf u}+\tilde{\bf z} ) -
 | {\bf z} + \sqrt{2\mu -1 }\alpha_{\bf u}\rangle
 \right\|+ \nonumber\\
 &&
 \left\|
U_{j} \left(\frac{\bf u}{\sqrt{n}}\right) U_{j} \left(\frac{\tilde{\bf z}}{\sqrt{n}}\right)
|jj \rangle  -
 U_{j} \left(\frac{{\bf u}+ \tilde{{\bf z}}}{\sqrt{n}}\right)   | jj \rangle
\right\|.
\label{eq.3terms}
\end{eqnarray}
This is obtained by adding and subtracting
$|\psi( n,j,\tilde{\bf z})\rangle \langle \psi( n,j,\tilde{\bf z})|$ and
$|\psi( n,j, {\bf u}+\tilde{\bf z})\rangle \langle \psi( n,j, {\bf u}+\tilde{\bf z})|$ and using the fact that $\| |\psi \rangle\langle \psi | - |\phi \rangle \langle\phi | \|_{1} = \sqrt{2}\|\psi-\phi\|$ for normalized vectors $\psi,\phi$.

The two first terms are similar, we want to dominate
them uniformly: we replace ${\bf u} + \tilde{\bf z}$ by $\tilde{\bf z}$ with
$|{\bf z}|\leq 2n^{\eta + \delta}$. We then write:
\begin{eqnarray}\label{eq.splitsum}
&& \left\|
\psi(n,j, \tilde{\bf z}) - |{\bf z}\rangle \right\|^{2}
=
\sum_{k=0}^{\infty} |\langle k| \psi(n,j, \tilde{\bf z})\rangle -\langle k|{\bf z}\rangle |^{2}\nonumber\\
&&\leq
\sum_{k=0}^{r-1} |\langle k| \psi(n,j, \tilde{\bf z})\rangle -\langle k|{\bf z}\rangle |^{2} +
2\sum_{k=r}^{\infty} \left(  |\langle k| \psi(n,j, \tilde{\bf z})\rangle |^{2} +|\langle k|{\bf z}\rangle |^{2}\right).
\end{eqnarray}
If ${\bf z} = |{\bf z}| e^{i\theta}$ then we have \cite{Hayashi&Matsumoto}
\begin{eqnarray*}
\langle k| \psi(n,j, \tilde{\bf z})\rangle
&=&
\sqrt{2j \choose k} \left(\sin(|{\bf z}|/\sqrt{n})e^{i\theta}\right)^k
\left(\cos(|{\bf z}|\sqrt{n})\right)^{2j-k},\\
\langle k|  {\bf z}\rangle &=&
\exp\left(-\frac{(2\mu-1)|{\bf z}|^2}{2}\right)\frac{\left(e^{i\theta}|{\bf z}|\sqrt{2\mu-1}\right)^k}{\sqrt{k!}}.
\end{eqnarray*}
%NEW: CHANGE IN THE DEFINITION OF r
In \eqref{eq.splitsum} we choose $r= n^{2\eta + \epsilon_3}$ with
$\epsilon_3$ satisfying the conditions
$ 2\delta+2\eta+\epsilon <2\eta+\epsilon_3 +\epsilon< 1/2 $
and $\eta+\epsilon_3 <1/4$. Then the tail sums are of the order
\begin{eqnarray*}
&&
\sum_{k=r}^{\infty} |\langle k | {\bf z} \rangle |^{2}
\leq \frac{ |{\bf z}|^{2 r} }{r !}\leq
\frac{(2n^{(\eta+\delta)})^{2 n^{2\eta + \epsilon_3}}}{(n^{2\eta
+\epsilon_3})!}=o\left(\exp(-n^{2\eta+\epsilon_3})\right),\\
&&
\sum_{k=r}^{\infty} |\langle k| \psi(n,j,\tilde{\bf z})\rangle |^{2} \leq
\sum_{k=r}^{j} \left( \frac{|{\bf z}|^{2}}{n} \right)^{k}
\frac{(2j)!}{(2j-k)! k!} \leq n \frac{|{\bf z}|^{2r}}{r!}
 = o\left(\exp(-n^{2\eta+\epsilon_3})\right).
\end{eqnarray*}
For the finite sums we use the following estimates which are uniform over
all $|{\bf z}|\leq 2n^{\eta+\delta}$, $k\leq r$, $j\in \mathcal{J}_{n}$:
\begin{align*}
\sqrt{2j \choose k} & =
\frac{((2\mu-1)n)^{k/2}}{\sqrt{k!}}(1+O(n^{-1/2+\epsilon +2\eta +
\epsilon_3})), \\
(\sin(|{\bf z}|/\sqrt{n}))^k
& = (|{\bf z}|/\sqrt{n})^k(1+O(n^{4 \eta +  \epsilon_3+ 2\delta -1} )) ,\\
(\cos(|{\bf z}|/\sqrt{n}))^{2j-k} &=
\exp\left(-\frac{(2\mu-1)|{\bf z}|^2}{2}\right) (1+ O(n^{2\eta-1/2 + \epsilon
+2\delta})),
\end{align*}
%NEW: EXPLICIT RATES ABOVE, EXPLANATIONS BELOW, FINAL RATE BY TAKING SQUARE
%ROOT
where we have used on the last line that $(1+x/n)^n = \exp(x) (1+
O(n^{-1/2}x))$ for $x\leq n^{1/2-\epsilon_4}$ (cf. \cite{Kahn}). This is enough
to show that the finite sum converges uniformly to zero at rate
$O(n^{2\eta-1/2 + \epsilon
+\epsilon_3})$ (the worst if $\epsilon_3$ is small enough) and
thus the first second terms in \eqref{eq.3terms} as the square root of this,
that is $O(n^{\eta-1/4+\epsilon/2 + \epsilon_3/2})$.

Notice that the errors terms depend on $\mu$ only through $j$, and that $2 j\geq
\epsilon n$ for $\mu-1/2\geq \epsilon_2$. Hence they are uniform in $\mu$.

%NEW: EXPLICIT RATES
We pass now to the third term of \eqref{eq.3terms}. By direct computation it can be shown that if we consider two general elements
$\exp(i  X_{1} )$ and $\exp(i X_{2})$ of $SU(2)$ with $X_{i}$ selfadjoint elements of $M(\mathbb{C}^{2})$
%$\vec{r} \vec{\sigma} := r_{x}\sigma_{x} + r_{y} \sigma_{y} + r_{z}\sigma_{z}$
then
\begin{equation}
\exp(-i(X_{1}+X_{2}))\exp(iX_{1}) \exp(iX_{2}) \exp([X_{1},X_{2}]/2) =
\mathbf{1} + O( X_{i_{1}} X_{i_{2}} X_{i_{3}}),
\end{equation}
where the $O(\cdot)$ contains only third order terms in $X_{1}, X_{2}$. If $X_{1},X_{2}$ are in the linear span of $\sigma_{x}$ and $\sigma_{y}$ then
all third order monomials are such linear combinations as well.

In particular we get that for ${\bf z},{\bf u}\leq n^{\eta+\epsilon_3}$:
\begin{eqnarray}
U(\beta):=&&
U\left(-\frac{{\bf u}+{\bf v}}{\sqrt{n}}\right)
U\left(\frac{\bf u}{\sqrt{n}}\right)U\left(\frac{\bf v}{\sqrt{n}} \right)
\exp(i (u_{x} v_{y}-u_{y}v_{x})\sigma_{z}/n)\nonumber
\\
=&& \begin{bmatrix}
1 + O(n^{-2+4\eta+4\epsilon_3}) & O(n^{-3/2 + 3\eta +3\epsilon_3}) \\
O(n^{-3/2 + 3\eta + 3\epsilon_3}) & 1 + O(n^{-2+ 4\eta +4\epsilon_3})
\end{bmatrix}\label{eq.curvature}.
\end{eqnarray}

Finally,using the fact that $|j,j\rangle$ is an eigenvector of $L_{z}$, the third term in \eqref{eq.3terms} can be written as
$$
\| |j,j\rangle\langle j,j| - U_j(\beta) |j,j\rangle\langle j,j|
U_j(\beta)^*\|
$$
and both states are pure, so it suffices to show that the scalar product converges to
to one uniformly. Using \eqref{eq.curvature} and the expression of
$\langle j| U_j (\beta)|j\rangle $ \cite{Hayashi&Matsumoto} we get, as $j\leq
n$,
$$
\langle j,j| U_j (\beta)|j,j\rangle = \left[ U(\beta)_{1,1}\right]^ j =1 + O(n^{-1
+ 4\eta + 4\epsilon_3}),
$$
which implies that the third term in \eqref{eq.3terms} is of order
$O(n^{-1+ 4\eta + 4\epsilon_3})$. By choosing $\epsilon_{3}$ and $\epsilon$
small enough, we obtain that all terms used in bounding (\ref{eq.gaussian.integral}) are uniformly
$O(n^{-1/4+ \eta + \epsilon  })$ for any $\epsilon>0$.

This ends the proof of
convergence \eqref{eq.channel.conv.} from the $n$ qubit state to the oscillator.

\subsection{ Proof of Theorem \ref{th.qlan}; the map $S_{n}$}\label{app.a2}

The opposite direction \eqref{eq.channel.conv.inverse} does not require much
additional estimation, so will only give an outline of the argument.

Given the state $N^{\bf u}\otimes \phi^{\bf u}$, we would like to map it into $\rho^{\bf u}_{n}$ or close to this state, by means of a completely positive map $S_{n}$.

Let $X$ be the classical random variable with probability distribution $N^{\bf u}$.
With $X$ we generate a random $j\in \mathbb{Z}$  as follows
$$
j(X) = [ \sqrt{n}X + n (\mu-1/2) ].
$$
This choice is evident from the scaling properties of the probability distribution
$p^{\bf u}_{n}$ which we want to reconstruct. Let $q^{\bf u}_{n}$ be the probability distribution of $j(X)$.
By classical local asymptotic normality results we have the convergence
\begin{equation} \label{wildzwijn}
\sup_{\|{\bf u} \|\leq n^{\eta}} \| q^{\bf u}_{n} - p^{\bf u}_{n}\|_{1} = O(n^{\eta-1/2}).
\end{equation}

Now, if the integer $j$ is in the interval $\mathcal{J}_{n}$ then
we prepare the $n$ qubits in block diagonal state with the only non-zero
block corresponding to the $j$'th irreducible representation of $SU(2)$:
$$
\tau^{\bf u}_{n,j} := \left(V^{*}_{j} \phi^{\bf u} V_{j} +
\mathrm{Tr}(P_{j}^{\perp} \phi^{\bf u}) \mathbf{1} \right)  \otimes \frac{\mathbf 1}{n_{j}}.
$$
The transformation $\phi^{\bf u}\mapsto \tau^{\bf u}_{n,j} $ is trace
preserving and completely positive \cite{Guta&Kahn}.

If $j\notin\mathcal{J}_{n}$ then we may prepare the qubits in an arbitrary state which
we also denote by $\tau^{\bf u}_{n,j}$.
The total channel $S_{n}$ then acts as follows
$$
S_{n}: N^{\bf u} \otimes \phi^{\bf u} \mapsto  \tau^{\bf  u}_{n} := \bigoplus_{j=0,1/2}^{n/2} \,
 q^{\bf u}_{n,j} \tau^{\bf u}_{n,j}.
$$

We estimate the error $\| \rho^{\bf u}_{n} -   \tau^{\bf  u}_{n} \|_{1}$ as
$$
\| \rho_{n}^{\bf u} -   \tau^{\bf  u}_{n} \|_{1} \leq
\| q^{\bf u}_{n} - p^{\bf u}_{n}\|_{1} +
2\mathbb{P}_{ p^{\bf u}_{n}} (j\notin\mathcal{J}_{n}) +
\sup_{j\in\mathcal{J}_{n}} \|\tau_{n,j}^{\bf u} - \rho_{n,j}^{\bf u}\|_{1}
$$
The first term on the r.h.s. is $O(n^{\eta - 1/2})$ (see \eqref{wildzwijn}),
the second term
is $O(n^{\epsilon - 1/2})$ (see \eqref{tail}). As for the third term, we use
the triangle inequality to write, for $j\in\mathcal{J}_{n}$,
$$
\| \tau_{n,j}^{\bf u} - \rho_{n,j}^{\bf u}\|_{1} \leq
\| \tau_{n,j}^{\bf u} - V_{j}^{*} \phi^{\bf u} V_{j}^{*}\|_{1} +
\| V_{j}^{*} \phi^{\bf u} V_{j}^{*} - \rho_{n,j}^{\bf u}\|_{1}\,.
$$
The first term is $O(e^{-n(1/2 - \eta - 2\delta)})$, according to the discussion
following equation \eqref{zwaardvis}.
The second term on the right is $O(n^{-1/4 + \eta + \epsilon})$
according to equations \eqref{eq.triangle.ineq.} through \eqref{eq.curvature}.

Summarizing, we have
$\|S_{n}(N^{\bf u} \otimes \phi^{\bf u}) - \rho^{\bf u}_{n} \|_{1}
= O(n^{-1/4 + \eta + \epsilon})
$, which establishes the proof in the inverse direction.

\qed

\section{Appendix: Proof of Theorem \ref{th.unitary.evolution}} \label{kool}

\noindent{\it First estimate.}
We build up the state $\tilde{\rho}^{\bf u}_{j,n}$ by taking linear
combinations of number states $|m\rangle$ to obtain an
approximate coherent state $|{\bf z}\rangle$, and finally mixing such states
with a Gaussian distribution to get an approximate displaced thermal state.
Consider the approximate coherent vector
$
P_{\tilde m} |{\bf z}\rangle ,
$
for some fixed ${\bf z}\in\mathbb{C}$ and $\tilde m = n^{\gamma}$, with $\gamma$ to be fixed later.
Define the normalized vector
\begin{equation}\label{eq.normalized.vectors}
|\psi^{n}_{{\bf z},j} \rangle : = \frac{1}{\| P_{\tilde m} |{\bf z}\rangle\| }
\sum_{m=0}^{\tilde m} \frac{|{\bf z}|^{m}}{\sqrt{m!}} |m\rangle,
\end{equation}
We mix the above states to obtain
$$
\tilde{\rho}^{\bf u}_{j,n} :=
\frac{1}{\sqrt{2\pi s^{2}}}\int e^{-|{\bf z -\sqrt{2\mu-1} \alpha_{\bf u}}|^{2}/2s^2}
\left(|\psi^{n}_{{\bf z},j} \rangle \langle \psi^{n}_{{\bf z},j}| \right)\,
d^{2}{\bf z}.
$$
Recall that $s^2 = (1-\mu)(4 \mu - 2)$, and
$$
\phi^{\bf u} = \frac{1}{\sqrt{2\pi s^{2}}}\int e^{-|{\bf z -\sqrt{2\mu-1} \alpha_{\bf u}}|^{2}/2s^2}
\left(|{\bf z} \rangle \langle {\bf z}| \right)\,
d^{2}{\bf z}.
$$
From the definition of $|\psi^{n}_{{\bf z},j} \rangle$ we have
\begin{equation}
\label{bound1}
\| |\psi^{n}_{{\bf z},j} \rangle- |{\bf z}\rangle\| \leq \sqrt{2} \frac{|{\bf z}|^{\tilde
m}}{\sqrt{\tilde m!}} \wedge 2,
\end{equation}
 which implies
$$
\| \tilde{\rho}^{\bf u}_{j,n} - \phi^{\bf u}\|_{1} \leq
 \frac{\sqrt{2}}{\sqrt{\pi s^{2}}}\int e^{-|{\bf z} |^{2}/2s^2}
\left( \frac{|{\bf z} + \sqrt{2\mu-1} \alpha_{\bf u}|^{\tilde m}}{\sqrt{\tilde m!}}\wedge \sqrt{2} \right)  \,
d^{2}{\bf z}= O(e^{-n^{2(\eta+\epsilon)}}),
$$
for any $\epsilon> 0$, for any $\gamma\geq 2(\eta + \epsilon)$.
Indeed we can split the integral into two parts. The integral over the domain
$|{\bf z}|\geq n^{\eta+\epsilon}$ is dominated by the Gaussian factor and
is $O(e^{-n^{2(\eta+\epsilon)}})$. The integral over the disk $|{\bf z}|\leq n^{\eta+\epsilon}$ is bounded by supremum of (\ref{bound1}) since the Gaussian integrates to one,
and is $O(e^{-(\gamma/2-\eta-\epsilon) n^{\gamma}})$. In the last step we use Stirling's formula to obtain
$\log \left[ (n^{\eta+\epsilon})^{n^{\gamma}} / \sqrt{n^{\gamma}!}\right] \approx  (\eta+\epsilon-\gamma/2) n^{\gamma} \log n$. Note that the estimate is uniform with respect
to $\mu-1/2> \epsilon_{2}$ for any fixed $\epsilon_{2}>0$.

\noindent
{\it Second estimate.} We now compare the evolved qubits
state $\tilde{\rho}^{\bf u}_{j,n}(t)$ and the evolved oscillator state $\phi^{\bf u}(t)$.
Let $\ket{\psi_{m,j}^{n}(t)} = U_{j,n}(t)\, |m\rangle \otimes \ket{\Omega}$ be the joint state at time $t$ when the initial state of the system is $|m\rangle$ corresponding to $|j, j-m\rangle$ in the $L_{z}$ basis notation. We choose the following approximation of $\ket{\psi_{m,j}^{n}(t)}$
\begin{equation}\label{eq.approx.m}
\ket{\xi_{m,j}^{n}(t)} :=\sum_{i=0}^{m} c_{n}(m,i) \alpha_{i}(t) | m-i\rangle \otimes
|e^{-1/2 u} \chi_{[0,t]}(u) \rangle_{i},
\end{equation}
where $\alpha_{i}(t) = \exp((-m+i)t/2)$,
$c_{n}(m,i) := c_{n}(m,i-1) \sqrt{\frac{2j-m+i}{2j_{n}}} \sqrt{\frac{m-i+1}{i}}$ with
$c_{n}(m,0):=1$, and $|f\rangle_{n} := f^{\otimes n}$ as defined in \eqref{eq.exponential}.
In particular for $\mu-1/2>\epsilon_2$ and $j\in\mathcal{J}_{n}$ we have
$c_{n}(m,i)\leq  \sqrt{\binom{m}{i} (1+
\frac{2}{\epsilon_2}n^{-1/2+\epsilon})^{i}}$.
% as long as $\gamma\leq 1/2+\epsilon$.

We apply now the estimate \eqref{eq.approx.qsde.solution}. By direct computations we get
\begin{eqnarray}
d\ket{\xi_{m,j}^{n}(t)}
&=&
 -\frac{1}{2}\sum_{i=0}^{m}
 c_{n}(m,i) \alpha_{i}(t)(m-i) |m-i \rangle \otimes |e^{-1/2 u} \chi_{[0,t]}(u) \rangle_{i} dt
\nonumber \\
&+&
 \sum_{i=1}^{m}  c_{n}(m,i) \alpha_{i - 1} (t) |m-i \rangle \otimes
 |e^{-1/2 u} \chi_{[0,t]}(u) \rangle_{i-1} \otimes_{s} |\chi_{[t,t+dt]}\rangle,
\label{eq.dxi}
\end{eqnarray}
 where
$$
f^{\otimes i} \,\otimes_{s}g := \sum_{k=1}^{i+1} f\otimes f \otimes \dots \otimes g\otimes \dots \otimes f.
$$
From the quantum stochastic differential equation we get
\begin{eqnarray}
&&
G_{dt} \, \ket{\xi_{m,j}^{n}(t)} =\nonumber\\
&&
-\frac{1}{2} \sum_{i=0}^{m} c_{n}(m,i) \alpha_{i}(t) (m-i) \frac{2j-m+i +1}{2j_{n}}  | m-i\rangle \otimes
|e^{-1/2 u} \chi_{[0,t]}(u) \rangle_{i} dt \label{eq.gdt}\\
&& + \sum_{i=0}^{m} c_{n}(m,i) \alpha_{i}(t) \sqrt{\frac{(m-i)(2j-m+i+1)}{2j_{n} (i+1)}}
 | m-i-1\rangle \otimes
|e^{-1/2 u} \chi_{[0,t]}(u) \rangle_{i} \otimes_{s} |\chi_{[t,t+dt]}\rangle.
 \nonumber
\end{eqnarray}
In the second term of the right side of \eqref{eq.gdt}  we can replace
$c_{n}(m,i) \sqrt{\frac{(m-i)(2j-m+i+1)}{2j_{n} (i+1)}}$ by $c_{n}(m,i+1)$ and
thus we obtain the same sum as in the second term of the left side of \eqref{eq.dxi}. Thus
\begin{eqnarray*}
&&
G_{dt} \ket{\xi_{m,j}^{n}(t)} - d \ket{\xi_{m,j}^{n}(t)} = \\
&&
\frac{1}{2}  \sum_{i=0}^{m-1} c_{n}(m,i) \alpha_{i}(t) (m-i)
\frac{2(j_{n}-j) +m-i-1}{2j_{n}} |m-i \rangle \otimes
 |e^{-1/2 u} \chi_{[0,t]}(u) \rangle_{i} \, dt.
\end{eqnarray*}
Then using  
$c_{n}(m,i)\leq  
\sqrt{\binom{m}{i}  (1+  (2/\epsilon_2) n^{-1/2 + \epsilon} )^{i}}
$ we get that $
\|G_{dt} \xi_{m,j}^{n}(t) - d\xi_{m,j}^{n}(t)\|$ is bounded from above by
\begin{eqnarray*}
&&
\frac{1}{2}\left[
\sum_{i=0}^{m-1}  \binom{m}{i}  ((1+ n^{-1/2+\epsilon})(1-e^{-t}))^{i} e^{-(m-i)t}
\left(\frac{(2(j_{n}-j) +m-i-1)(m-i)}{2j_{n}} \right)^{2} \right]^{1/2}dt.
\end{eqnarray*}
 We have
$$
\frac{(2(j_{n}-j) +m-i-1)(m-i)}{2j_{n}} = O( m( n^{-1/2 +\epsilon} + n^{-1}m))
$$
Inside the sum we recognize the binomial terms with the $m$'th term missing.
Thus the sum is
\begin{eqnarray*}
&&
\left(1+ n^{-1/2+\epsilon} -e^{-t} n^{-1/2+\epsilon}  \right)^{m} -
\left((1-e^{-t}) (1+n^{-1/2+\epsilon})\right)^{m} \\
&&
\leq
(1+n^{-1/2+\epsilon})^{m} ( 1 - (1-e^{-t})^{m} )\leq
(1+n^{-1/2+\epsilon})^{m} \, m e^{-t}.
\end{eqnarray*}
Then there exists a constant $C$ (independent of $\mu$ if $\mu-1/2\geq
\epsilon_2$) such that
$$
\|G_{dt} \xi_{m,j}^{n}(t) - d\xi_{m,j}^{n}(t)\| \leq  \frac{C}{2} e^{-t/2}
 m^{3/2} (n^{-1/2+\epsilon} +m n^{-1})   \left(1+\frac{2}{\epsilon_{2}} n^{-1/2+\epsilon}\right)^{m/2}
$$
By integrating over $t$ we finally obtain
\begin{equation}\label{eq.approx.error.m}
\|\psi_{m,j}^{n}(t)- \xi_{m,j}^{n}(t)\| \leq  C
 m^{3/2} (n^{-1/2+\epsilon} +m n^{-1})   \left(1+ \frac{2}{\epsilon_{2}}n^{-1/2+\epsilon}\right)^{m/2}  .
\end{equation}
Note that under the assumption $\gamma < 1/3-2\epsilon/3$, the right side converges to zero at rate $n^{3\gamma/2- 1/2+\epsilon}$ for all $m\leq\tilde{m}= n^{\gamma}$. Summarizing, the assumptions which we have made so far over $\gamma$ are
$$
2\eta+2\epsilon <\gamma<1/3-2\epsilon/3.
$$

Now consider the vector
$
\ket{\psi^{n}_{{\bf z},j}}% :=P_{\tilde m} |{\bf z}\rangle ,
$
as defined in \eqref{eq.normalized.vectors} and let us denote
$\ket{\psi_{{\bf z},j}^{n}(t)}= U_{j,n}(t) \ket{\psi^{n}_{{\bf z},j} } \otimes \ket{\Omega}$.
Then based on \eqref{eq.approx.m} we choose the approximate solution
$$
\ket{\xi_{{\bf z},j}^{n}(t)}= e^{-|{\bf z}|^{2}/2} \sum_{m=0}^{\tilde m} \frac{|{\bf
z}|^{m}}{\sqrt{m!}}
\sum_{i=0}^{m} c_{n}(m,i) \alpha_{i}(t) | m-i\rangle \otimes
|e^{-1/2 u} \chi_{[0,t]}(u) \rangle_{i}.
$$
Note that the vectors $\ket{\psi_{k,j}^{n}(t)}$ and $\ket{\xi_{k,j}^{n}(t)}$ live in the ``$k$-particle'' subspace of $\mathcal{H}_{j}\otimes \mathcal{F}(L^{2}(\mathbb{R}))$ and thus are orthogonal to all vectors $\ket{\psi_{p,j}^{n}(t)}$ and $\ket{\xi_{p,j}^{n}(t)}$ with $p\neq k$.
By \eqref{eq.approx.error.m},  the error is
\begin{eqnarray}
\|\psi_{{\bf z},j}^{n}(t) - \xi_{{\bf z},j}^{n}(t) \|
&\leq&
C e^{-|{\bf z}|^{2}/2}
\left( \sum_{m=0}^{\tilde m} \frac{|{\bf z}|^{2m}}{m!}
 m^{3} (n^{-1/2+\epsilon} +m n^{-1})^{2}   \left(1+ \frac{2}{\epsilon_{2}}n^{-1/2+\epsilon}\right)^{m}\right)^{1/2}
 \nonumber\\
 &+& \frac{|{\bf z}|^{2\tilde{m}}}{\tilde{m}!}\nonumber
\\
 &\leq&
C\tilde{m}^{3/2}  (n^{-1/2+\epsilon} +\tilde{m} n^{-1})   \left(1+ \frac{2}{\epsilon_{2}}n^{-1/2+\epsilon}\right)^{\tilde m/2}
+ \frac{|{\bf z}|^{2\tilde{m}}}{\tilde{m}!}.\label{eq.xi-psi.j}
\end{eqnarray}

We now compare the approximate solution $\xi_{{\bf z} ,j}^{n}(t)$ with the ``limit'' solution
$\psi_{{\bf z}}(t)$ for the oscillator coupled with the field as described in section \ref{sec.oscillator.field}. We can write
$$
\psi_{{\bf z}}(t) = e^{-|{\bf z}|^{2}/2} \sum_{m=0}^{\infty} \frac{|{\bf z}|^{m}}{\sqrt{m!}}
\sum_{i=0}^{m} \sqrt{\binom{m}{i}} e^{-(m-i)t/2} | m-i\rangle \otimes
|e^{-1/2 u} \chi_{[0,t]}(u) \rangle_{i}.
$$
Then
\begin{eqnarray*}
&&
\| \xi_{{\bf z},j}^{n}(t) - \psi_{\bf z}(t) \|^{2}  =\\
&&
e^{-|{\bf z}|^{2}}  \sum_{m=0}^{\tilde m} \frac{|{\bf z}|^{2m}}{m!} \sum_{i=0}^{m}
e^{-(m-i)t} \left|c_{n}(m,i) - \sqrt{\binom{m}{i}} \right|^{2} (1-e^{-t})^{i} +
e^{-|{\bf z}|^{2}} \sum_{m=\tilde m}^{\infty} \frac{|{\bf z}|^{2m}}{m!} .
\end{eqnarray*}

Now
\begin{eqnarray*}
\left|c_{n}(m,i) -\sqrt{ \binom{m}{i}} \right|^{2}
&\leq&
\left| c_{n}(m,i)^{2} - \binom{m}{i}\right|\\
&\leq&
 \binom{m}{i}
\left|1- \prod_{p=1}^{i} \left(1+ \frac{ 2(j- j_n) - m +p }{2j_{n}} \right)\right|
\\
&\leq&
C_{2} \binom{m}{i}  m n^{-1/2+\epsilon},
\end{eqnarray*}
where $C_2$ does not depend on $\mu$ as long as $\mu-1/2\geq\epsilon_2$ (recall
that the dependence in $\mu$ is hidden in $j_n = (2\mu-1)n$). Thus
\begin{equation}\label{eq.xi-psi}
\| \xi_{{\bf z},j}^{n}(t) - \psi_{\bf z}(t) \|^{2}\leq
C_{2} n^{-1/2+\epsilon}
e^{-|{\bf z}|^{2}}  \sum_{m=0}^{\tilde m} \frac{ m|{\bf z}|^{2m}}{m!}  +  \frac{|{\bf z}|^{2\tilde m}}{\tilde m!} \leq
C_{2} n^{-1/2+\epsilon} |{\bf z}|^{2} + \frac{|{\bf z}|^{2\tilde m}}{\tilde m!} .
\end{equation}

From \eqref{eq.xi-psi.j} and \eqref{eq.xi-psi} we get
\begin{eqnarray*}
&&
\|\psi_{{\bf z},j}^{n} (t) -  \psi_{\bf z}(t)\| \leq \\
&&
2 \wedge \left[C\tilde{m}^{3/2}  (n^{-1/2+\epsilon} +\tilde{m} n^{-1})
\left(1+ \frac{2}{\epsilon_{2}}n^{-1/2+\epsilon}\right)^{\tilde m/2} + \frac{|{\bf z}|^{2\tilde{m}}}{\tilde{m}!} +
 \left[C_{2} n^{-1/2+\epsilon} |{\bf z}|^{2} + \frac{|{\bf z}|^{2\tilde
m}}{\tilde m!}   \right]^{1/2} \right] \\
&&
:= E(\tilde{m}, n, {\bf z})
\end{eqnarray*}

We now integrate the coherent states over the displacements ${\bf z}$ as we did in the case of local asymptotic normality in order to obtain the thermal states in which we are interested
$$
\tilde{\rho}^{\bf u}_{j,n} :=
\frac{1}{\sqrt{2\pi s^{2}}}\int e^{-|{\bf z -\sqrt{2\mu-1} \alpha_{\bf u}}|^{2} /2s^{2}}
\left(|\psi^{n}_{{\bf z},j} \rangle \langle \psi^{n}_{{\bf z},j}| \right)\,
d^{2}{\bf z}.
$$

We define the evolved states
$$
\tilde{\rho}^{\bf u}_{j,n} (t) := U_{j,n}(t) \tilde{\rho}^{\bf u}_{j,n} U_{j,n}(t)^{*},
\qquad
\mathrm{and}
\qquad
\phi^{\bf u} (t):= U(t) \phi^{\bf u} U(t)^{*},
$$
Then
$$
\sup_{j\in\mathcal{J}_{n}} \sup_{\|{\bf u}\|\leq  n^{\eta}} \| \tilde{\rho}^{\bf u}_{j,n} (t) - \phi^{\bf u} (t)\|_{1} \leq \sup_{\|{\bf u}\|\leq  n^{\eta}}
 \frac{1}{\sqrt{\pi s^{2}}}\int e^{-|{\bf z -\sqrt{2\mu-1} \alpha_{\bf u}}|^{2}/2s^{2}}
 E(\tilde{m}, n, {\bf z})\,
d^{2}{\bf z}.
$$

Here again we cut the integral in two parts. On $|{\bf z}|\geq n^{\eta +
\epsilon}$, the Gaussian dominates, and this outer part is less
than $e^{-n^{\eta+\epsilon}}$. Now the inner part is dominated by $\sup_{|{\bf
z}|\leq n^{\eta + \epsilon}} E(\tilde{m},n,{\bf z})$. Now we want $\tilde m$ to
be not too big for (\ref{eq.xi-psi.j}) to be small, on the other hand, we want
${\bf z}^{2\tilde m}/ {\tilde m}!$ to go to zero. A choice which satisfies the condition
is $\gamma = 2\eta + 3\epsilon$. By renaming $\epsilon$ we then get
\[
 E(\tilde{m}, n, {\bf z}) = O(n^{\eta-1/4+ \epsilon }, n^{3\eta -1/2 +\epsilon}),
\]
for any small enough $\epsilon>0$.
Hence we obtain \eqref{eq.unitary.error}.

\qed

%\bibliography{bib_osfld}

\end{document}